\documentclass[twocolumn]{aastex62}

\usepackage{savesym}
\usepackage{mathtools}
\savesymbol{tablenum}
\usepackage{multirow}
\usepackage[natbib]{}
\usepackage{graphicx}

\graphicspath{{./}{figures/}}

\shorttitle{CO-Galaxy Cross-correlation}
\shortauthors{Keenan et al.}

\def\xunits{$\mu$K~h$^{-3}\,$Mpc$^3$}
\def\aunits{$\mu$K$^2$~h$^{-3}\,$Mpc$^3$}

\def\pxtotal{$P_\mathrm{\times}=-4\pm270$ \xunits{}}
\def\pxlimit{$P_\mathrm{\times}<540$ \xunits{}}

\def\lumlimittp{$L^\prime<3.3\times10^{10}$ K km s$^{-1}$ pc$^2$}
\def\lumlimitfit{$L^\prime<4.4\times10^{10}$ K km s$^{-1}$ pc$^2$}
\def\lumstack{$L^\prime=(-0.5\pm2.5)\times10^{10}$ K km s$^{-1}$ pc$^2$}
\def\lumlimstack{$L^\prime<5.0\times10^{10}$ K km s$^{-1}$ pc$^2$}

\def\pxfitnpnum{$-482\pm357$}
\def\pxfitnp{$P_\mathrm{\times}=-482\pm357$ \xunits{}}

\def\bbTfitnpnum{$122\pm60$}
\def\bbTfitcoldznum{$63\pm45$}
\def\bbTfitnp{$b_\mathrm{gal}b_\mathrm{CO}T_\mathrm{CO}=122\pm 60$ $\mu$K}
\def\bbTfitcoldz{$b_\mathrm{gal}b_\mathrm{CO}T_\mathrm{CO}=63\pm 45$ $\mu$K}

\def\Tvalnum{$8.4\pm6.0$ $\mu$K}

\def\Tlim{$\langle T_{\rm CO}\rangle < 20.4$ $\mu$K}

\def\rholim{$\rho_\mathrm{H_2}<1.5\times10^9$ M$_\odot$ Mpc$^{-3}$}

\defcitealias{keating+15}{K15}
\defcitealias{keating+16}{K16}
\defcitealias{keating+20}{K20}

\begin{document}

\title{An Intensity Mapping Constraint on the CO-Galaxy Cross Power Spectrum at Redshift $\sim3$}

\correspondingauthor{R. P. Keenan}
\author[0000-0003-1859-9640]{Ryan P. Keenan}
\altaffiliation{NSF Graduate Research Fellow}
\affiliation{Steward Observatory, University of Arizona, 933 North Cherry Avenue, Tucson, AZ 85721, USA}
\email{rpkeenan@email.arizona.edu}

\author[0000-0002-3490-146X]{Garrett K. Keating}
\affiliation{Center for Astrophysics, Harvard \& Smithsonian, 60 Garden Street, Cambridge, MA 02138, USA}

\author[0000-0002-2367-1080]{Daniel P. Marrone}
\affiliation{Steward Observatory, University of Arizona, 933 North Cherry Avenue, Tucson, AZ 85721, USA}

\begin{abstract}

The abundance of cold molecular gas plays a crucial role in models of galaxy evolution. While deep spectroscopic surveys of CO emission lines have been a primary tool for measuring this abundance, the difficulty of these observations has motivated alternative approaches to studying molecular gas content. One technique, line intensity mapping, seeks to constrain the average molecular gas properties of large samples of individually undetectable galaxies through the CO brightness power spectrum. Here we present constraints on the cross-power spectrum between CO intensity maps and optical galaxy catalogs. This cross-measurement allows us to check for systematic problems in CO intensity mapping data, and validate the data analysis used the auto-power spectrum measurement of the CO Power Spectrum Survey. We place a $2\sigma$ upper limit on the band-averaged CO-galaxy cross-power of \pxlimit{}. Our measurement favors a non-zero $\langle T_{\rm CO}\rangle$ at around 90\% confidence and gives an upper limit on the mean molecular gas density at $z\sim2.6$ of $7.7\times10^8$ M$_\odot$Mpc$^{-3}$. We forecast the expected cross-power spectrum by applying a number of literature prescriptions for the CO luminosity to halo mass relation to a suite of mock light cones. Under the most optimistic forecasts the cross-spectrum could be detected with only moderate extensions of the data used here, while more conservative models could be detected with a factor of 10 increase in sensitivity. Ongoing CO intensity mapping experiments will target fields allowing extensive cross correlation analysis and should reach the sensitivity required to detect the cross-spectrum signal.

\end{abstract}

\keywords{}

\section{Introduction} \label{sec:intro}

Cold molecular gas represents the raw material for star formation. Understanding its abundance over cosmic time is a necessary ingredient for theories of galaxy formation and evolution \citep{carilli+13,walter+20}. Blind surveys of emission from CO, the preferred tracer of the total molecular gas content in galaxies, have begun to probe its redshift evolution \citep{lenkic+20,decarli+19,pavesi+18,decarli+16,walter+14}. However, until at least the advent of the ngVLA, these types of survey will be limited to sky areas of a few square arcminutes and only tens of secure direct detections \citep{decarli+20}. As a result, these samples are subject to large statistical uncertainties which make it difficult to determine trends with redshift \citep{keenan+20}.

Line intensity mapping (LIM) offers a complementary approach to direct detection efforts. Instead of searching for individual objects at high significance, LIM measures line intensity in large fields. The  integrated luminosity of every line-emitting galaxy is extracted from the power spectrum of the three-dimensional line intensity distribution and used to constrain the properties of galaxies too faint to be individually detected \citep{visbal+10,lidz+11,gong+11,breysse+14,li+16}. Such data sets can be used to constrain the luminosity function of CO emission lines and thus the abundance evolution of molecular gas. The first line intensity mapping measurements have provided constraints on the total CO abundance at redshifts $1<z<6$ \citep{pullen+13,keating+15,keating+16,uzgil+19,keating+20}. With dedicated intensity mapping instruments now taking data (e.g. TIME; \citealt{sun+21}, COMAP; \citealt{ihle+19}, CONCERTO; \citealt{ade+20}), this technique is set to advance our understanding of high redshift molecular gas in the near future.

However, understanding the co-evolution of star formation and molecular gas abundance will require synthesis of radio and submillimeter gas measurements with data from optical and IR (OIR) galaxy surveys. Placing the statistical results of intensity mapping in the context of individually detected OIR galaxies represents a potential challenge. One promising path forward is through cross correlation between catalogs from galaxy surveys and CO maps produced by LIM experiments  \citep{wolz+16,wolz+17,chung+19a}. By making use of the large-scale clustering present in the maps and the galaxy distribution, cross-correlation can extract information beyond what is derived from stacking spectra extracted at known galaxy positions. Cross-correlations between intensity maps and a broad range of other data sets have been explored as a tool for addressing numerous astrophysical and cosmological problems, including calibration of photometric redshifts \citep{cunnington+19}, characterizing feedback from active galactic nucleii \citep{breysse+19}, determining physical properties of the interstellar medium \citep[ISM;][]{sun+19}, exploring the Ly$\alpha$ forest \citep{carucci+17}, measuring baryon acoustic oscillations at high redshift \citep{cohn+16}, mitigating the effects of cosmic variance \citep{oxholm+21_arxiv}, and constraining models of primordial non-Gaussianity \citep{dizgah+19}, among others.

Cross-correlation also serves as a check on systematics in intensity mapping data, and can be used to validate detections of the auto-correlation power spectrum \citep{furlanetto+07,silva+15}. This is true not only for CO but also ongoing hydrogen 21cm and [C II]~158$\mu$m intensity mapping projects \citep{chang+10,masui+13,wolz+21_arxiv}. 

The CO Power Spectrum Survey \citep[COPSS;][hereafter K15 and K16]{keating+15,keating+16} was a first-generation intensity mapping experiment, which obtained thousands of hours of observations targeting emission from the CO(1-0) transition at redshift $z \sim 2.3$ to $3.3$, constraining the total luminosity of CO during the peak of cosmic star formation. The COPSS observations were designed to allow cross-correlation by targeting regions with extensive coverage in optical/infrared spectroscopy. Here we study the cross-correlation between deep 30 GHz observations of the GOODS-N field from COPSS and a large catalog of galaxies with spectroscopic redshifts.

In Section~\ref{sec:formalism} we review the mathematical formalism of the LIM technique and cross-correlation. In Section~\ref{sec:obs} we describe the 30 GHz and optical data we use. We detail our data analysis procedure for both cross-correlation and stacking in Section~\ref{sec:analysis} and present the results of this analysis in Section~\ref{sec:results}. In Section~\ref{sec:validation} we describe tests to verify that our analysis is not unduly affected by systematics. In Section~\ref{sec:model} we model the CO-galaxy cross spectrum in the presence of measurement errors, to verify our analysis methodology and quantify measurement uncertainties. We place our results in the context of theoretical expectations and upcoming experiments, and use them to constrain the average CO luminosity and cosmic molecular gas density at $z\sim3$ in Section~\ref{sec:discussion}. We conclude in Section~\ref{sec:conclusion}. Throughout we assume a flat $\Lambda$CDM cosmology with $H=70$ km s$^{-1}$ Mpc$^{-1}$ and $\Omega_M=0.27$.

\section{Intensity Mapping Formalism} \label{sec:formalism}

The power spectrum of an intensity field measures the variance in brightness temperature fluctuations as a function of spatial frequency. It encodes information about the total luminosity of line emitting sources as well as information about their clustering. Here we present the formalism in terms of the CO(1-0) line transition, but note that it applies to any emission line. The power spectrum can be expressed as
\begin{equation}\label{eq:pco}
    P_{\rm CO}(k,z) = \langle T_{\rm CO}\rangle^2 b_{\rm CO}(z)^2 P_{\rm lin}(k,z) + P_{\rm shot,CO}(z) ,
\end{equation}
where $P_{\rm CO}(k,z)$ is the power spectrum at comoving scale $k$ and redshift $z$, $\langle T_{\rm CO}\rangle$ is the average brightness temperature of CO emission in the volume under consideration, $b_{\rm CO}$ is the bias of CO emitters, $P_{\rm lin}$ is the linear matter power spectrum, and $P_{\rm shot,CO}$ is the shot noise due to the discrete nature of individual tracers. 

The shot term can be further written as 
\begin{equation}\label{eq:pshot}
    P_{\rm shot,CO} = \Big( \frac{c^3(1+z)^2}{8\pi k_B H(z)\nu_{\rm rest}^3} \Big)^2 \int_0^\infty L_{\rm CO}^2 \frac{dn}{dL} dL ,
\end{equation} 
where $c$ is the speed of light, $k_B$ is the Boltzmann constant, $H(z)$ is the Hubble parameter at redshift $z$, $\nu_{\rm rest}$ is the rest frequency of CO(1-0) emission, 115.27 GHz, $L_{\rm CO}$ is CO luminosity, and $dn/dL$ is the luminosity function for CO emitters. The terms preceding the integral serve to convert from luminosity to brightness temperature. On scales smaller than a few Mpc ($k\ga1$~Mpc$^{-1}$), the shot power is the dominant contribution to the power spectrum while the clustering power dominates at large scales.

The cross power-spectrum of a galaxy overdensity field and a CO intensity field may be written as 
\begin{equation}\label{eq:px}
    P_{\rm \times}(k,z) = \langle T_{\rm CO}\rangle b_{\rm CO}(z) b_{\rm gal}(z) P_{\rm lin}(k,z) + P_{\rm shot,\times}(z) ,
\end{equation}
where $b_{\rm gal}$ is the bias of the galaxies in the selected galaxy survey. 

The cross-shot power $P_{\rm shot,\times}$ is
\begin{equation}\label{eq:pxshot}
    P_{\rm shot,\times} = \frac{1}{n_{\rm gal}} \Big( \frac{c^3(1+z)^2}{8\pi k_B H(z)\nu_{\rm rest}^3} \Big) \sum_{\rm cat} \frac{L_{\rm CO,i}}{V} ,
\end{equation}
where $n_{\rm gal}$ is the number density of catalog galaxies, $L_{\rm CO,i}$ is the CO luminosity of the ith member of the galaxy catalog, $V$ is the volume sampled, and the sum is taken over the galaxies in the catalog \citep{breysse+19}. The cross-shot power can thus be used to measure the average CO luminosity of the optical galaxy catalog, equivalent to measuring the average CO luminosity of catalog galaxies via stacking.

\section{Observations and Data} \label{sec:obs}

\subsection{CO Observations}

The CO observations used here were conducted with the Sunyaev-Zel'dovich Array (SZA), an 8-element array of 3.5 m antennas, part of the Combined Array for Research in Millimeter-wave Astronomy (CARMA). Observations were conducted from 2013 April to 2015 April as part of COPSS. The fields of the COPSS survey were observed over a frequency range of 27 to 35 GHz, covered in 16 spectral windows of 500 MHz bandwidth. The field of view of the SZA is $\sim$13 arcminutes at the low frequency end of the band and $\sim$10 arcminutes at the high frequency end. Details of the survey can be found in \citetalias{keating+16}. We use data for the GOODS-N field from both the pilot and primary phases of the COPSS survey.

\subsection{Galaxy Catalogs}

A number of large spectroscopic surveys have targeted GOODS-N. We have attempted to compile the most complete set of spectroscopic redshifts available by synthesizing these into a single catalog. We draw redshifts from the following sources:
\begin{itemize}
    \item The MOSFIRE Deep Evolution Field (MOSDEF) survey of NIR selected galaxies at $z\ga1.4$ using the Keck MOSFIRE spectrograph \citep{kriek+15}, giving 124 galaxies.
    \item The Team Keck Redshift Survey 2 (TKRS2) with Keck/MOSFIRE \citep{wirth+15}, with 19 galaxies.
    \item A sample 23 of MOIRCS Deep Survey BzK galaxy redshifts from \citet{yoshikawa+10}.
    \item The catalog of \citet{barger+08} which compiles most prior redshifts as well as new measurements and includes a total of 2710 redshifts.
    \item A survey of optically selected BM/BX and LBG galaxies conducted with the Keck LRIS-B spectrograph \citep[Reddy catalog;][]{reddy+06,steidel+04}.
\end{itemize}

The same galaxy may appear in multiple catalogs. Therefore, we search for objects within 1 arcsecond (about twice the typical MOSDEF seeing) and $\Delta z/(1+z) < 0.003$ in two or more of the catalogs and remove objects from the older data set, we also remove objects with offsets $<0.25$ arcsecond irrespective of redshift. 

We then cut the catalog to the redshift range covered by the COPSS frequency band. For CO(1-0) emission this range is $2.3<z<3.2$.

These cuts leave us with a total of 124 MOSDEF galaxies, 17 additional galaxies from TKRS2, 0 galaxies for MOIRCS, 78 galaxies from \citet{barger+08} and 3 galaxies from \citet{reddy+06} for a total of 224 galaxies. Sixteen of these fall in gaps between SZA spectral windows, leaving 208 galaxies. 

\begin{figure*}
    \centering
    \includegraphics[width=.8\textwidth]{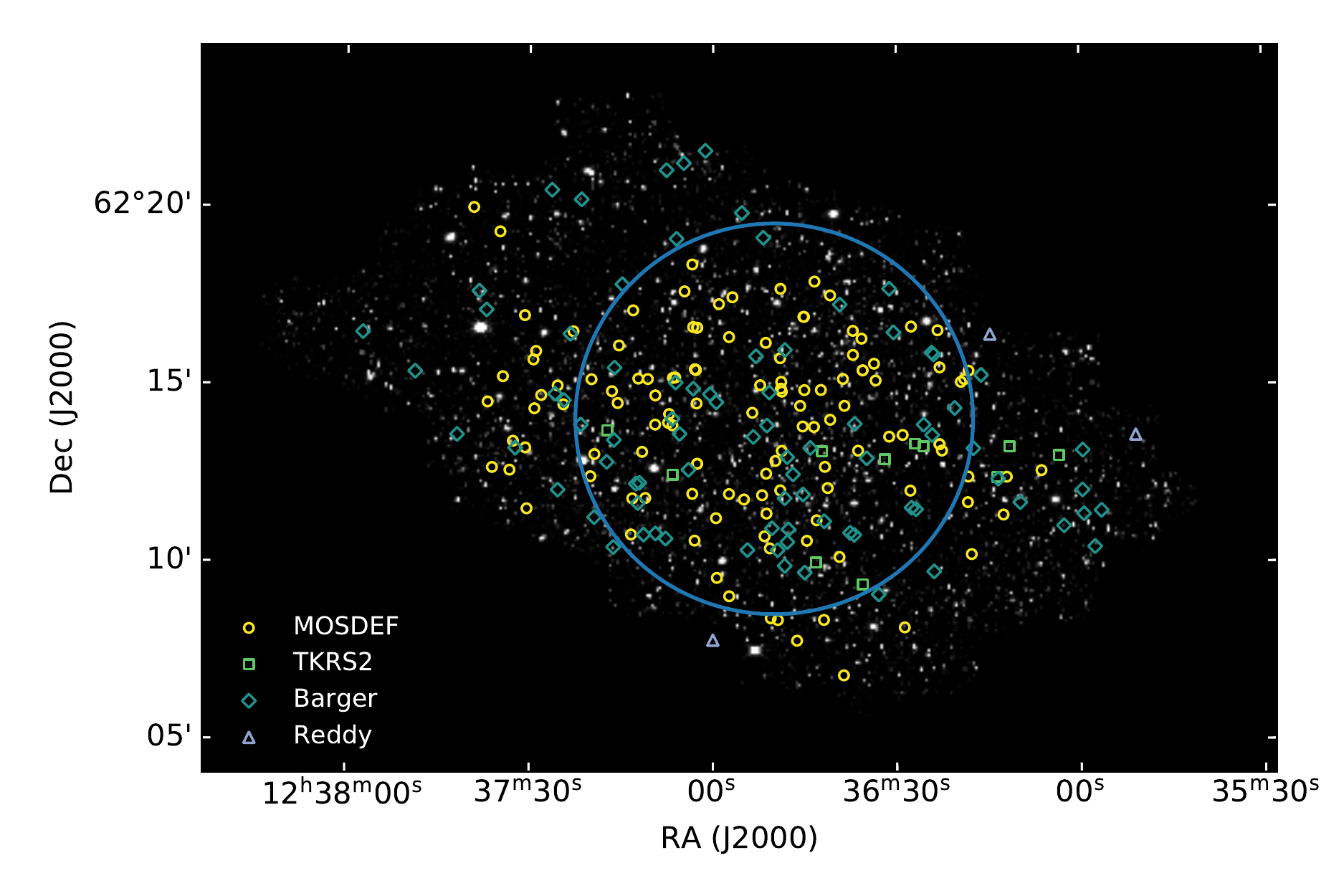}
    \caption{The sample of $2.3<z<3.2$ galaxies with spectroscopic redshifts, ploted over the 3dHST combined F140W+F125W+F160W image of GOODS-N. Galaxies drawn from MOSDEF, TKRS2, \citet{barger+08}, and \citet{reddy+06}, are indicated with yellow circles, green squares, blue diamonds, and lavender triangles respectively. The large blue circle is the 11 arcminute primary beam of the SZA at 31 GHz.}
    \label{fig:map}
\end{figure*}

\begin{figure}
    \centering
    \includegraphics[width=.45\textwidth]{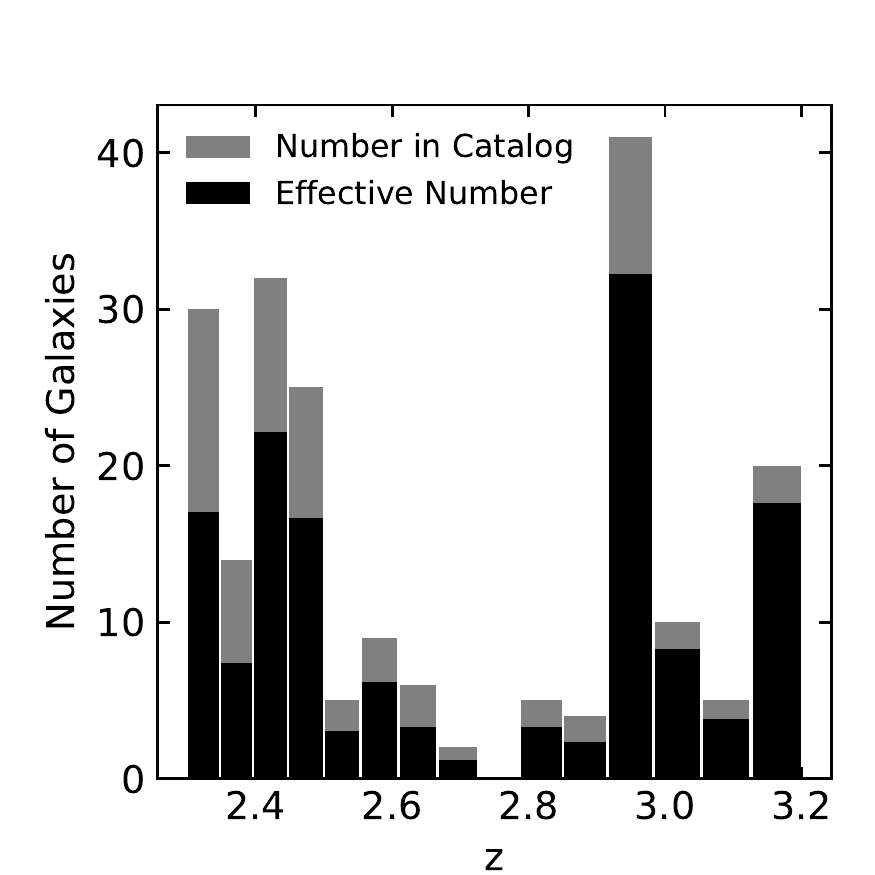}
    \caption{The redshift distribution of our final galaxy sample. The bins correspond to frequency windows of our SZA data. Shown in grey are the total number of galaxies in each window. Due to the tapering of the SZA primary beam, not every galaxy in the field received equal weight in our analysis. We define the effective number of galaxies, with each galaxy weighted by the primary beam response at its position. The black histogram shows this distribution. As the primary beam becomes larger at lower frequency (higher redshift) while the redshift catalog area remains fixed, the effective number of galaxies increases relative to the total in high redshift windows.}
    \label{fig:zdist}
\end{figure}

The positions of all galaxies in our data sample are shown in Figure~\ref{fig:map}, and their redshift distribution within the spectral windows of our SZA data is shown in Figure~\ref{fig:zdist}. Because the primary beam of the telescope tapers from the pointing center, we also count galaxies weighted by the primary beam of the SZA, which gives us an effective total of 145 galaxies.

In addition to spectroscopic catalogs, extensive grism spectroscopy from the 3D-HST program \citep{momcheva+16} is available. However, \citet{chung+19a} find that the redshift uncertainty of low resolution of grism spectra results in significant attenuation of the cross power, and therefore we do not include them. This results in the exclusion of 69 grism redshifts.

Each of the catalogs discussed above is selected on the basis of different properties and for different science goals. Therefore, they do not represent a homogeneous population of galaxies. For the purpose of maximizing the sensitivity of our CO-galaxy cross-power spectrum we nevertheless use the combined data sets. We also explored cross-correlating the individual catalogs with the COPSS data, but did not detect the cross-power for any subset of galaxies explored. Cross-correlation of subsets of the galaxy population will have to await more sensitive intensity mapping data sets.

\section{Data Analysis} \label{sec:analysis}

The cleaning and analysis pipeline for the SZA data is described in detail in \citetalias{keating+15} and \citetalias{keating+16}. We provide a brief summary here. The raw interferometric data are recorded as three dimensional arrays gridded in angular frequency along the sky plane (u and v) and in frequency ($\nu$). During the primary COPSS survey, trailing fields offset by five and ten minutes in right ascension were observed at the same declination as the GOODS-N field, in order to facilitate removal of contamination from antenna cross-talk and ground emission. Data from all three fields are averaged to produce a model of the contamination, and the model is subtracted from each field. This is not possible for data from the pilot phase of the survey, which were taken without trailing fields, and contamination is a significant problem in short baselines. To reduce this contamination, measurements exceeding $4\sigma$ significance are removed.

In auto-correlation, ground contamination will be a source of excess power and must be carefully removed from the data to avoid a positive bias. In cross-correlation, the ground contamination does not correlate with galaxy positions, and therefore naturally drops out of the cross-spectrum without introducing any bias. However, random alignments of the galaxies and ground emission introduce additional noise.  For our data, we find that cleaning the ground-contamination improves the sensitivity of cross-spectrum despite the noise that is added as a result of field differencing, and we retain this step. 

The grids are next Fourier transformed along the frequency axis within each spectral window, to produce ``delay-visibilities". For each window this procedure produces a grid in (u, v, $\eta$) where $\eta$ is the Fourier transform of frequency, also called the delay.

We expect continuum sources to be spectrally smooth over the frequency range of each window. For typical spectral indices seen at 30 GHz, these sources will predominantly show power in the $\eta=0$ bin. To remove these sources we drop this $\eta$ bin from our grids. \citetalias{keating+16} verify that this procedure produces auto-power spectra without significant contamination from continuum sources.

Next, the grid is converted from observable (u, v, $\eta$) coordinates to comoving Mpc units at the redshift of the expected CO signal:
\begin{equation}\label{eq:k}
\mathbf{k}=(k_x,k_y,k_z)=(\frac{2\pi u}{X(z)}, \frac{2\pi v}{X(z)}, \frac{2\pi\eta}{Y(z,\nu_{\rm rest})})
\end{equation}
where $X=D_M(z)$ the comoving transverse distance at redshift $z$ and $Y=c(1+z)^2/H(z)\nu_{\rm rest}$, $c$ is the speed of light, $H(z)$ is the reshift-dependent Hubble parameter, and $\nu_{\rm rest}$ is the rest frequency of the target line (115.27 GHz for CO(1-0)).

\subsection{CO Auto Spectra}\label{ss:auto}

To produce the CO auto-power spectrum, we use the estimator
\begin{equation}
    \mathcal{P}(\mathbf{k})=\frac{1}{V_{\rm eff}}\Big[\frac{\Sigma_\mathbf{k^\prime} \sigma_k^{-2}\sigma_{k^\prime}^{-2} C(\mathbf{k-k^\prime})(|\tilde{T}^*(\mathbf{k})\tilde{T}(\mathbf{k^\prime})|)}{\Sigma_\mathbf{k^\prime} \sigma_k^{-2}\sigma_{k^\prime}^{-2} C(\mathbf{k-k^\prime})} - \mathcal{A}_\mathbf{k}\Big]
\end{equation}
\begin{equation}\label{eq:spherical}
    P(k)=\langle \mathcal{P}(\mathbf{k}) \rangle_{\mathbf{k\cdot k}=k^2}
\end{equation}
where $\mathcal{P}$ is the three-dimensional power spectrum, $\sigma_\mathbf{k}$ is the estimated thermal noise, $C$ is the expected normalized covariance matrix, and $\tilde{T}$ is the delay visibility. $V_{\rm eff}$ is the effective volume probed by the measurement, which for a Gaussian beam of solid angle $\Omega$ and bandwidth $B$ is given by $V_{\rm eff}=X^2YB\Omega/2$. $\mathcal{A}$ is the sum of the autocorrelations of individual delay visibilities within each grid cell, which is subtracted to remove noise bias. The averaging in Equation~\ref{eq:spherical} is done weighting each cell by the inverse of its estimated variance due to thermal noise level.

\subsection{Cross Spectra}\label{ss:cross}

To compute the cross-power spectrum we construct a grid in angular coordinates ($l$, $m$) and frequency ($\nu$) to capture the galaxy distribution. To calculate the latter, we compute the redshifted frequency of the CO(1-0) line given the reported optical redshift of the galaxy, such that $\nu=\nu_{\rm rest}/(1+z)$. We populate the grid by counting the number $N$ of galaxies falling within each cell. Then we compute the overdensity field $\delta_N(l,m,\nu)=(N-\bar N)/\bar N$ where $\bar N$ is the mean number of galaxies per cell in the region covered by the optical surveys. The resultant grid is then Fourier transformed across all three dimensions and converted into $\mathbf{k}$ coordinates via Equation~\ref{eq:k}.
 
The finite resolution of the grid results in quantization errors, which will cause in some decorrelation between the optical and radio data sets. To mitigate this, we use a grid three times finer in angular resolution, and five times finer in frequency resolution than the SZA data cubes. 

To estimate the cross-power spectrum, we use the estimator
\begin{equation}\label{eq:3dx}
    \mathcal{P}_\times(\mathbf{k}) = \frac{1}{V_{\times,{\rm eff}}}\mathcal{R}\Big(\frac{\tilde{T}^*(\mathbf{k})\tilde{\delta_N}(\mathbf{k})}{A(\mathbf{k})}\Big)
\end{equation}
\begin{equation}\label{eq:sphericalx}
    P_\times(k)=\langle \mathcal{P}_\times(\mathbf{k}) \rangle_{\mathbf{k\cdot k}=k^2}
\end{equation}
where $V_{\times,{\rm eff}}$ is the effective volume of the cross power measurement, $\mathcal{R}$ denotes the real part of a complex number, $\tilde{T}$ and $\tilde{\delta_N}$ are the Fourier duals of $T$ and $\delta_N$, and $A(k)$ corrects for attenuation of the observed power spectrum (this correction is detailed Section~\ref{ss:atten}).

To determine the effective volume we define a survey footprint function $F(l,m)$ to be 1 at positions $(l,m)$ within the footprint of the optical galaxy catalogs of GOODS-N and 0 elsewhere. This footprint is determined primarily by HST imaging coverage of the field \citep{skelton+14}, with slight expansions to cover a small number of galaxies from \citet{reddy+06} that lie beyond the edges of the HST coverage. We then define 
\begin{equation}\label{eq:Veff}
    V_{\times,{\rm eff}} = \int F(l,m)\times W(l,m) d\Omega
\end{equation}
where $W(l,m)$ is the primary beam response of the SZA.

The averaging in Equation~\ref{eq:sphericalx} is done with inverse variance weighting of each grid cell, after applying attenuation corrections. These weights differ from those for the auto-power spectrum because the galaxy field does not contribute additional thermal noise when multiplied with the CO delay visibilities. 

To estimate the noise in the cross power spectrum of each window, we randomized the phases of the SZA delay visibilities and recomputed the cross power using the same galaxy grid. The phase randomization removes all position information from the radio data sets, so this is equivalent to computing the cross-spectrum of the galaxy field with random noise in the radio data set. We perform $10^4$ trials to generate the expected noise distribution for each frequency window. We then determine the $1\sigma$ noise level from the 15.9 and 84.1 percentiles of the distribution.

We construct final power spectra by combining each spectral window and determine the average power by further averaging across all $k$-modes. Our cross-power spectrum results are reported in Section~\ref{sec:results} and Figure~\ref{fig:cross}.

\subsubsection{Attenuation Correction}\label{ss:atten}

In practice, a number of observational effects are known to attenuate the observed auto- and cross-spectra. Errors in spectroscopic redshifts can shift galaxies between cells of our overdensity field, causing position mismatches with the CO temperature field that cause decorrelation along the line of sight for small scale modes of the cross-spectrum \citep{chung+19a}. Moreover, galaxy rotation and bulk motions of the ISM broaden CO emission lines, resulting in typical line profiles hundreds of km s$^{-1}$ wide. As these line widths are comparable to the size of the spectral channels in our SZA data, we expect the signal to spread into adjacent channels. This will result in attenuation of both the auto- and cross-spectra along the line of sight \citep{chung+21_arxiv}. 

The attenuation of the COPSS auto-power spectrum has been explored in \citet{keating+20}. This effect is found to be fairly small, and as we make no effort to interpret our auto-spectrum here, we do not consider attenuation corrections to the auto-power estimator from Section~\ref{ss:auto}. However, the additional effect of redshift errors can significantly attenuate the cross-spectrum signal, particularly at high $k$. Therefore we attempt to correct this attenuation in our three-dimensional cross-power spectrum.

The attenuation of the cross-spectrum can be expressed as 
\begin{equation}\label{eq:atten}
    A(\mathbf{k}) = \exp\Big[-\frac{1}{2}k_z^2 \times (\frac{\sigma_\mathrm{lw}^2}{c^2} + \sigma_z^2) \Big(Y(z,\nu_\mathrm{rest})\frac{\nu_\mathrm{rest}}{1+z}\Big)^2\Big]
\end{equation}
where $k_z$ is the line of sight component of $\mathbf{k}$, $\sigma_\mathrm{lw}$ describes the width of the CO emission lines and for a Gaussian line profile is equal to the full width at half maximum (FWHM) divided by 2.355, and $\sigma_z$ is the redshift uncertainty. The final term puts these widths of these uncertainties in units of comoving length. 

\citet{kriek+15} compare the spectroscopic redshifts from the MOSDEF survey and a compilation of prior spectroscopic measurements, finding a typical $\sigma_z/(1+z)$ of $0.001$, which we take as our estimate of this term.

\citet{chung+21_arxiv} find that using a single, characteristic line profile is adequate for accounting for the effect of line width in many applications. We therefore estimate the typical FWHM of CO emitters in our data cube to be $\sim 300$ km s$^{-1}$. This value is broadly consistent with the median line width of 360 km s$^{-1}$ from a compilation of blindly detected CO lines from the ASPECS \citep{gonzalez-lopez+19} and COLDz \citep{pavesi+18} deep fields in the redshift range $1.0\lesssim z\lesssim 3.6$.

In Section~\ref{sec:model} we revisit our choices in modeling line width and redshift uncertainty, finding that across the range of likely values, the uncertainty introduced to the final power spectrum is negligible in comparison to the instrumental noise, and smaller than the effects other uncertainties such as cosmic variance.

\subsection{Stacking Analysis}

The shot power in the CO-galaxy cross-spectrum is proportional to the mean CO luminosity of the galaxies in the catalog. This mean luminosity can also be measured from our CO data cube via stacking. To facilitate comparison between these two approaches, we construct a stack on the positions of our catalog galaxies.

To perform this analysis we image the GOODS-N data using natural weighting. The resulting image has a typical 1$\sigma$ sensitivity of 0.27 Jy km s$^{-1}$ at the center of the beam and a synthesized beam width of 1.6 arcminutes. We then extract a cutout from the primary beam-corrected image at the position and expected frequency of each source. We use only galaxies with expected CO(1-0) line frequencies at least 31.25 MHz (one channel) from the edge of a spectral window and within 15 arcminutes of the map center. We convert each extracted snapshot to units of luminosity per spectral channel using
\begin{equation}
    L^\prime_\mathrm{chan}=3.25\times10^7 S\Delta v_\mathrm{chan}\frac{D_L(z)^2}{(1+z)^3\nu_\mathrm{obs}^2}\ \mathrm{K\ km\ s}^{-1}\ \mathrm{pc}^2
\end{equation}
where $S$ is the channel flux density in Jy, $\Delta v_\mathrm{chan}$ is the velocity width of the channel in km s$^{-1}$, $D_L$ is the luminosity distance in Mpc, $z$ is the redshift of the source, and $\nu_\mathrm{chan}$ is the frequency of the channel. The channel width of the COPSS data is 31.25 MHz, which corresponds to a velocity widths between 270 and 340 km s$^{-1}$. We then compute the inverse variance weighted sum of the extracted spectra for all of our cutouts. 

A redshift uncertainty of $\Delta z/(1+z)\sim0.001$, appropriate for the spectroscopic redshifts in our catalog, corresponds to a shift in expected CO line frequency of approximately one channel. CO line widths correlate with the line luminosity, and typically do not exceed 1,000 km s$^{-1}$ even for the most luminous high redshift galaxies \citep{carilli+13}. Most galaxies within our stack will have line widths well below this limit. Therefore we expect that most of the flux will fall within the three central channels of our stack. We therefore sum over these three channels to compute the luminosity of the stack.

\section{Results} \label{sec:results}

We find an auto-power of $P_{CO}=1700 \pm 3500$ \aunits{}, corresponding to a $2\sigma$ upper limit of $P_{\rm CO}<8700$ \aunits{} for the GOODS-N field. This result is consistent with the measurement of \citetalias{keating+16} using the full COPSS data set of $P_{CO}=3000 \pm 1300$ \aunits{}. We note that the ``GOODS-N'' result reported in \citetalias{keating+16} also includes data from the trailing fields, which are not included in our analysis because they have no corresponding optical data. As our auto-power results are not constraining, we do not consider them further in this paper. When constructing a joint constraint from the auto- and cross-power spectra we instead use the the final COPSS auto-power spectrum from \citetalias{keating+16}.

\begin{figure}
    \centering
    \includegraphics[width=.45\textwidth]{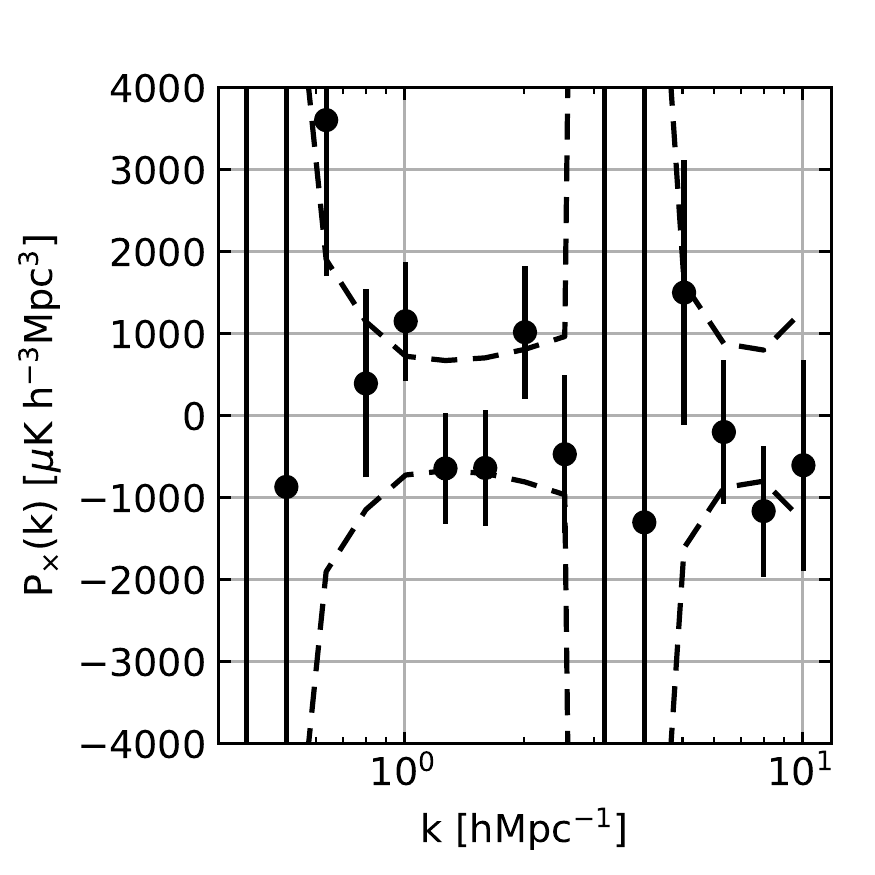}
    \caption{The results of our cross-spectrum analysis between the COPSS 30~GHz observations of GOODS-N and our spectroscopic catalog of OIR galaxies. The dashed curves show the $1\sigma$ sensitivity level of our measurement.}
    \label{fig:cross}
\end{figure}

Figure~\ref{fig:cross} shows our computed cross-power spectrum for the GOODS-N field. Taking the inverse variance weighted average over all $k$, we find a cross-power of \pxtotal{}. This corresponds to a $2\sigma$ upper limit of \pxlimit{}.

\begin{figure}
    \centering
    \includegraphics[width=.45\textwidth]{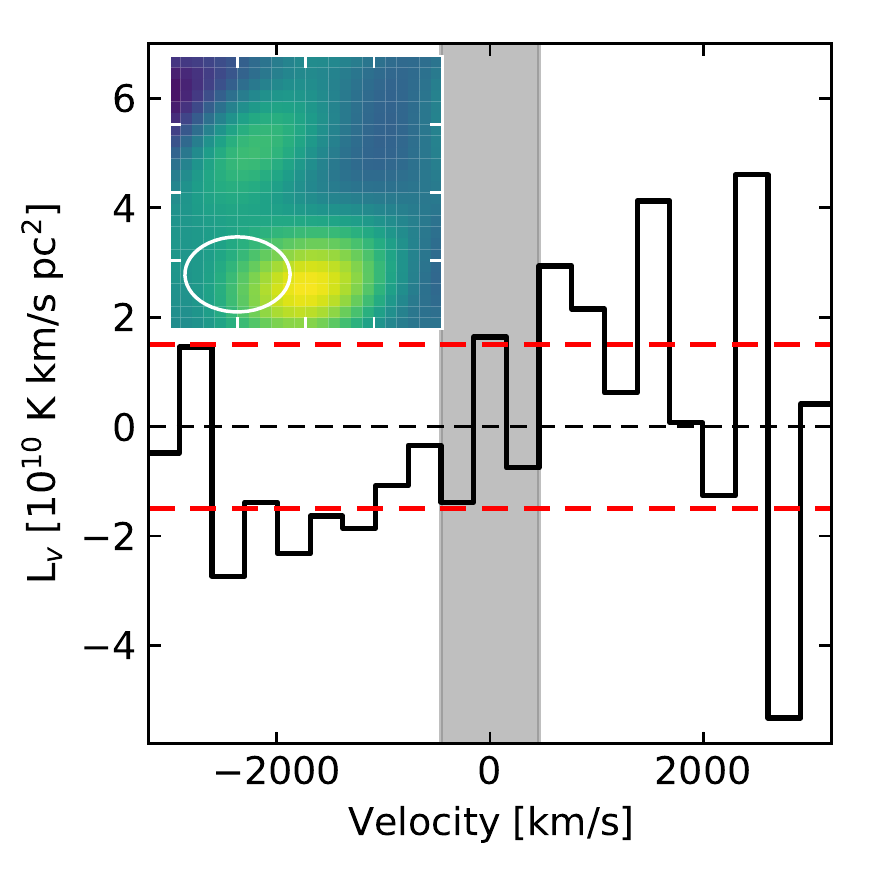}
    \caption{The average spectrum of our GOODS-N galaxy catalog, produced by stacking at the expected frequency of the CO(1-0) emission line. The red dashed lines indicate the average 1$\sigma$ noise level in a channel (which matches the channel-by-channel uncertainty within a few percent), while the grey shaded region shows the three channels integrated to obtain our constraint on the average CO luminosity luminosity. Inset: $4^\prime \times 4^\prime$ image of the central three channels of our stack. The spectrum is extracted from the central pixel of the map. The synthesized beam is shown in the lower left corner.}
    \label{fig:stack}
\end{figure}

Figure~\ref{fig:stack} shows the stacked spectrum of our galaxy catalog. We find no evidence of a detection of CO emission in the stack, finding \lumstack{} corresponding to an upper limit of \lumlimstack{}.

\section{Data Validation}\label{sec:validation}

\begin{figure*}
    \centering
    \includegraphics[width=\textwidth]{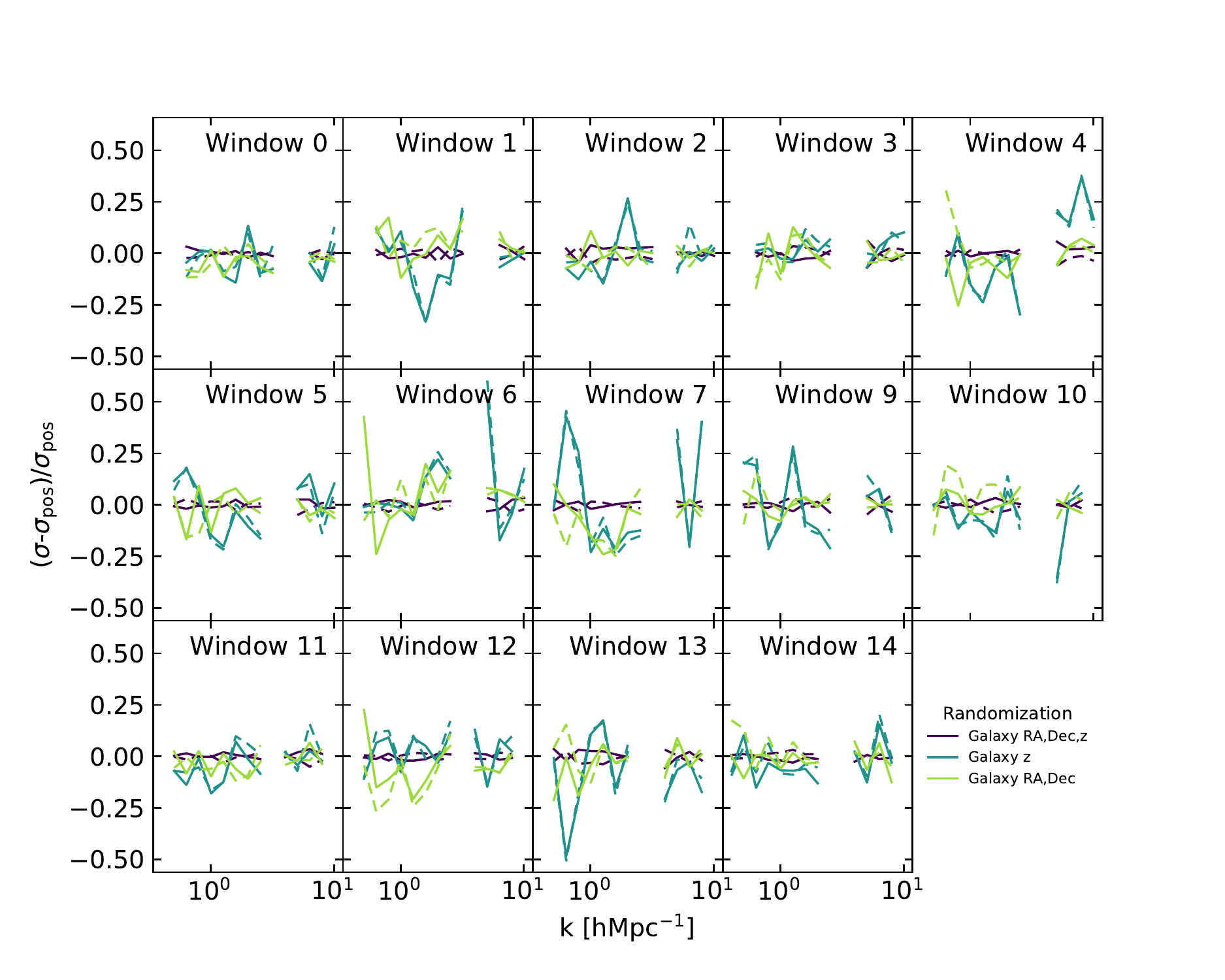}
    \caption{The plus (solid) and minus (dashed) $1\sigma$ noise curves for our galaxy position (dark blue), redshift (blue-green), and position angle (green) randomizations normalized to the average $1\sigma$ curve of the position randomization. Results of the three randomizations match, indicating no excess of power due to continuum sources which would appear in the randomizations for frequency but not position or position angle.}
    \label{fig:randomizations}
\end{figure*}

Optical and radio observations are subject to very different systematics. Radio intensity mapping measurements can be corrupted by terrestrial and astronomical foregrounds, but these are unrelated to the optical/IR galaxy catalogs used in the cross-correlation measurement. We therefore expect that contamination problems seen in the auto-power spectrum will be greatly diminished in the cross-spectrum. Here we search for evidence of contamination in our cross-spectra and compare these results to known contamination issues in the auto-spectra.

\subsection{Continuum Source Contamination}

We perform a number of additional randomizations as a check on noise and systematics. First, we randomize the phase of the galaxy grid and compute the cross-power with the un-randomized radio data. We verify that this produces results consistent with the radio phase randomizations, finding $1\sigma$ noise curves from the galaxy phase randomizations match within a few percent.

Next we randomize the redshifts (or equivalently frequencies) of galaxies within each window. This serves as a check of our continuum removal as galaxies matched in angular position to a continuum source will remain matched when shifted in frequency, producing excess low-$k$ power. For comparison, we produce a set of randomizations where each galaxy is assigned a new angular position within the footprint of the galaxy survey, in addition to a new redshift. Figure~\ref{fig:randomizations} shows the ratio of the $\pm1\sigma$ noise curves for redshift randomization and the full three-dimensional position randomization. The curves generally match, and we find no indication that the redshift-only randomization shows an excess relative to the other two tests. We therefore conclude that the contribution from continuum sources to our measurement is negligible.

\subsection{Mitigating Ground Contamination}

\begin{figure}
    \centering
    \includegraphics[width=.45\textwidth]{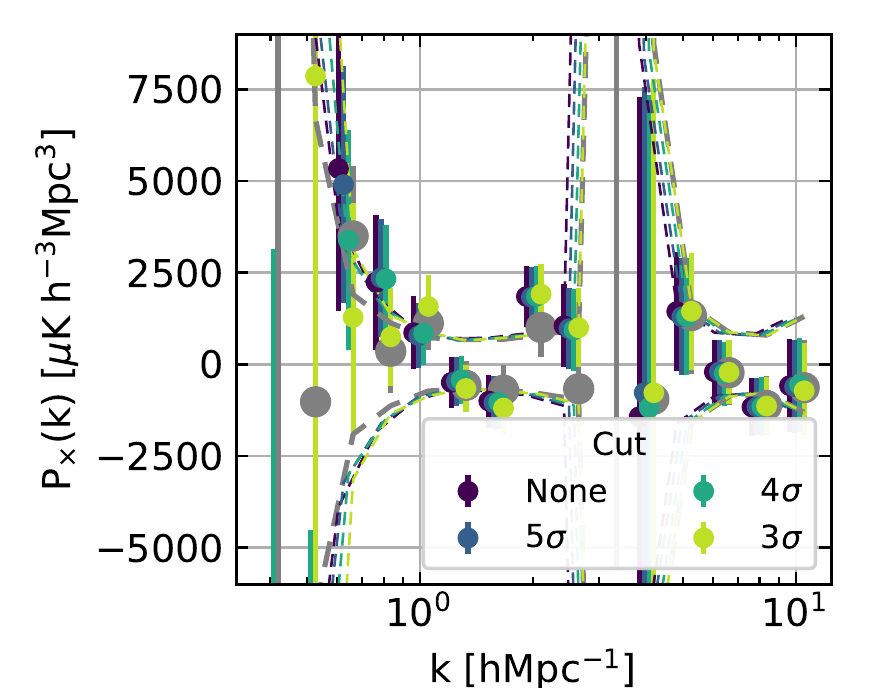}
    \caption{The cross-power spectrum for GOODS-N, reproduced using a range of different significance cuts to clean ground-correlated emission from the COPSS data. Values for each cut are offset slightly in k for clarity. The gray points show our final power spectrum, produced with a 4$\sigma$ cut along with trailing field subtraction. The cross-power levels are consistent for all spectra shown.}
    \label{fig:cuts}
\end{figure}

In our reduction of the SZA data we used observations from trailing fields to model and remove ground-correlated emission, and when this was not possible flagged baselines containing any measurement exceeding $4\sigma$ significance. To test the effect of the contaminated modes on the cross-signal, we re-ran our analysis without the trailing field subtraction, using only flagging at a range of significance cuts, including an analysis where no data are excluded. Figure~\ref{fig:cuts} shows the cross-power spectrum of the GOODS-N data for each cut. 

Interestingly, the sensitivity curves for the data with no trailing field subtraction fall slightly above those with the subtraction at low $k$, even though differencing fields increases the thermal noise in the CO data. This suggests that, while not the dominant source, systematics can contribute a non-negligible level of noise relative to the thermal noise. 

Even without any corrections, there is no significant excess power in the cross spectrum. The results are relatively insensitive to the significance cut applied, and the power spectrum with no corrections is consistent within errors of that constructed with both trailing field subtraction and a significance cut. By contrast, performing auto-power spectrum calculations without correcting for ground contamination produces power at many times the expected noise level, which swamps any astrophysical signal (see Appendix~\ref{appendix:auto_contamination} and \citetalias{keating+15}). This demonstrates the utility of cross-correlation for cleaning contaminated data; we could use the data in cross-correlation analyses, without introducing biases and with only a moderate increase in the noise level of the resulting power spectra, even if no correction for ground contamination was possible.

\subsection{Search for Interloper Line Contamination}\label{ss:foregrounds}

Emission lines from other molecules in galaxies at lower redshift can appear in intensity mapping data cubes and be confused for emission from the target line. This emission contributes to the auto-power spectrum in a manner that cannot easily be disentangled from the power of the target line \citep{cheng+16, cheng+20}. However, the sources of these lines will not appear in the galaxy density cubes used in cross-correlation, and therefore these interlopers will not contribute to the cross-power spectrum. In addition, if galaxy catalogs are available at the redshift of the confusing sources, they can be used to measure the contributions of those sources and assess the level at which they contribute to the auto-power.

At the redshifts studied by COPSS, no significant interlopers are expected for the CO(1-0) transition \citep{chung+17}, although dense gas tracers such as HCN J=1-0 and CS J=1-0 have been proposed as a potential contaminant \citep{breysse+15}. HCN emission in our SZA data corresponds to sources at $1.5<z<2.3$. Our galaxy catalogs contain 227 galaxies in this redshift range. We cross-correlate these galaxies with the SZA data in the same manner as Section~\ref{ss:cross} and find results consistent with zero. This allows us to place a $2\sigma$ upper limit on the total HCN cross-power of $P_{\times,HCN}<566$ \xunits{}.

CS J=1-0 emission would appear in the SZA bandpass at $0.4<z<0.8$. In this redshift range, our galaxy catalogs are considerably more complete. We cross-correlate the SZA data with a sample of 835 galaxies from our GOODS-N catalog. The measured power is again consistent with zero, and we place a $2\sigma$ upper limit on the shot noise in the CS cross-power spectrum of $P_{\times,CS}<44$ \xunits{}. 

In practice, contamination of the auto-power spectrum by low redshift interlopers is expected to be primarily driven by a handful of the brightest objects, while our cross-power limits reflect the average luminosity of all objects in our catalog. We therefore cannot rule out the possibility that a few extremely luminous objects are present in our field. However, models where contamination is expected to be significant require objects with HCN luminosities $\gtrsim10^{10}$ K km s$^{-1}$ pc$^2$ \citep{chung+17}. Even if such galaxies do exist, they would be so rare as to be unlikely to appear in our survey volume, therefore it seems unlikely that they contaminate the auto-power results of \citetalias{keating+16} significantly. This is further confirmed by analysis of the COPSS image cubes, which show no evidence for bright emission lines.

\section{Modeling}\label{sec:model}

Here, we use mock observations to better understand any systematics in our data analysis procedure. In Section~\ref{ss:moddescription} we describe a fiducial model of the galaxy-CO cross-power spectrum. Then in Section~\ref{ss:systematics} we assess the degree of attenuation caused by redshift uncertainties, spectral line emission being spread over multiple frequency channels, and the SZA's sampling of the $uv$-plane. We use these results to validate the attenuation corrections applied to our observed three dimensional power spectrum. In Section~\ref{ss:modcomparisons} we extend this analysis to literature models of the CO power spectrum in order to contextualize the sensitivity of our data.

\subsection{Model Description}\label{ss:moddescription}

We follow the approach of \citet[][hereafter L16; see also \citealt{silva+15,chung+17}]{li+16}, assigning CO luminosities to halo catalogs from dark matter simulations using a series of scaling relations. This approach is particularly advantageous for modeling cross correlation, as each CO emitter can be matched to other halo properties in order to construct mock galaxy catalogs for cross-correlation. We use the scaling relation between halo mass and CO luminosity fit to the COPSS auto-power spectrum by \citet{keating+20} as our fiducial model:
\begin{equation}\label{eq:k20model}
    L_\mathrm{CO}=
    \begin{cases*}
        A_\mathrm{CO}\frac{M^2}{M_0} & $M\leq M_0$ \\
        A_\mathrm{CO}M_0 & $M>M_0$
    \end{cases*}
\end{equation}
where $M$ is the halo mass, $A_\mathrm{CO}$ is the mass to CO luminosity ratio, and $M_0$ is a cutoff mass, above which the CO luminosity of halos remains approximately constant. $M_0$ is set to be $10^{12}$ h$^{-1}$M$_\odot$, and the fit to the COPSS auto power spectrum yields $A=1.6\times10^{-6}$ L$_\odot$ M$_\odot^{-1}$. To account for galaxy-to-galaxy variations, a log-normal scatter with $\sigma_{\rm CO}=0.37$ dex is added to the $M_{\rm halo}$-$L_{\rm CO}$ scaling for each halo. We implement this model using galaxy catalogs from the publicly available IllustrisTNG-300 simulation \citep{TNG1,TNG2,TNG3,TNG4,TNG5}.

To simulate observations of our models, we draw random lines of sight through the simulation box, selecting galaxies within 40 arcminutes of the field center to include in light cones, and allowing redshift evolution by stepping through simulation snapshots. The procedure for generating light cones is described in detail in \citealt{keenan+20}. The light cones are subdivided into frequency and angular resolution elements matched to the resolution of our data and for each element a brightness temperature is calculated. The light cones are then apodized according to the primary beam response of the SZA antennas.

Next, we select a catalog of galaxies to utilize for cross-correlation. To approximately reproduce the galaxy distribution in our catalog, we group spectral windows into three broad ranges: two high density bins from $2.3\la z\la 2.6$ and $2.8\la z \la 3.2$ and a low density bin in the $2.6 \la z \la 2.8$ gap. The high density bins correspond to the redshift ranges in which [O III] emission falls in a NIR window and can be used to make secure redshift determinations in IR spectroscopic surveys such as MOSDEF \citep{kriek+15}. The redshift ranges covered in \citet{reddy+06} also match approximately to this selection. Within each bin, we calculate the total number of galaxies in our real redshift catalog and require that our mock catalogs have this number of galaxies in the matching redshift range. We then populate our galaxy catalogs by drawing halos with $M>10^{11}$ M$_\odot$, which roughly corresponds to the $10^9$ M$_\odot$ stellar mass limit of the MOSDEF survey. MOSDEF prioritizes high mass targets, and so we weight our selection similarly.

We use these catalogs to construct galaxy density fields using the right ascension, declination, and observed redshift in a manner analogous to Section~\ref{sec:analysis}.

\subsection{Evaluation of Signal Attenuation and Systematics}\label{ss:systematics}

\begin{figure}
    \centering
    \includegraphics[width=.45\textwidth]{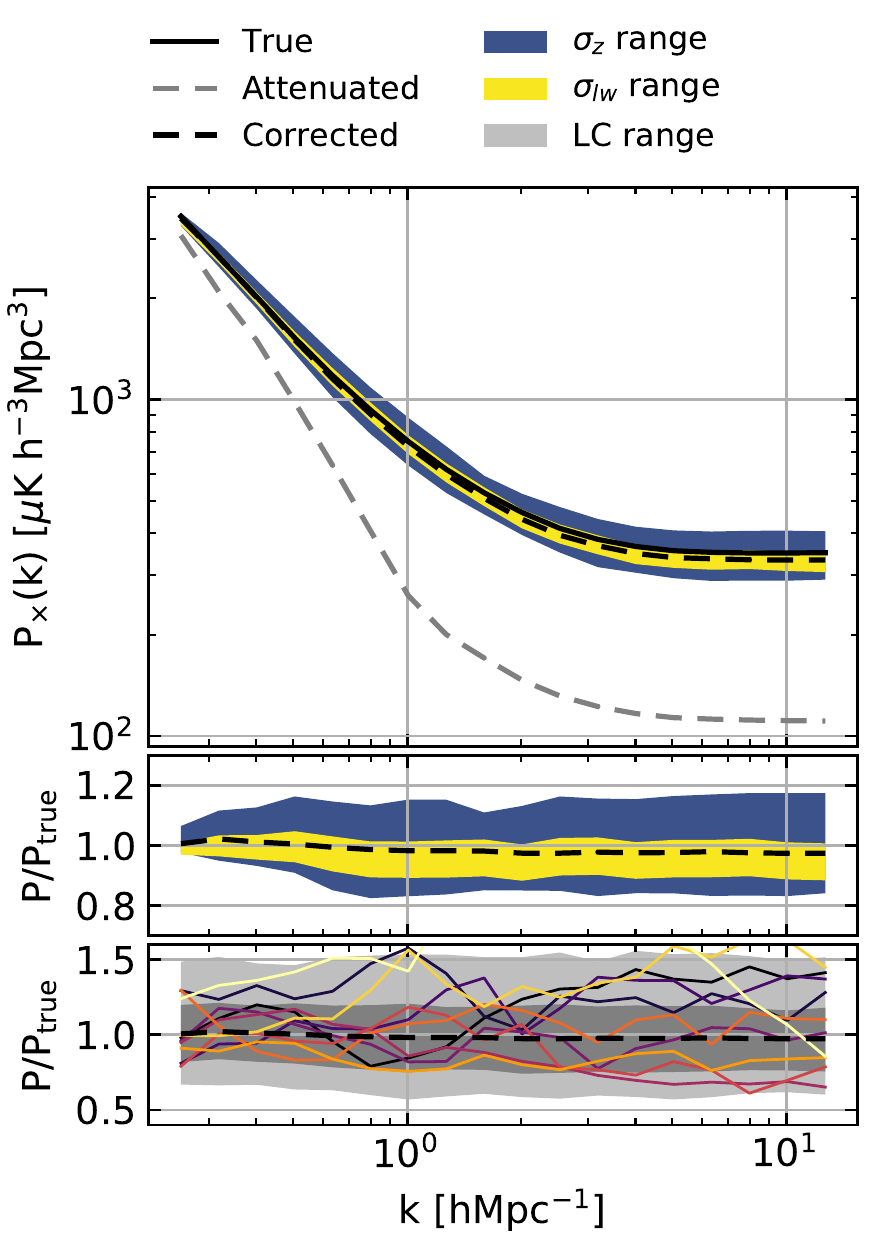}
    \caption{Top: The mean cross power spectra of our simulated light cones. The solid black line shows the true spectrum with no attenuation. The gray dashed lines shows the power spectrum attenuated by redshift errors of $\sigma_z/(1+z)=0.001$ and Gaussian line profiles with 300 km s$^{-1}$ FWHM, and the black dashed line shows this same power spectrum after applying the attenuation corrections described in Section~\ref{ss:atten}. The yellow region shows the effect of assuming a 300 km s$^{-1}$ line width when the true width ranges from 100-500 km s$^{-1}$, while the blue region shows the effect of assuming $\sigma_z/(1+z)=0.001$ when the true value ranges from 0.0007 to 0.0013.
    Middle: The ratios of attenuation corrected spectra to the true power spectrum, showing the degree of uncertainty introduced by incorrect assumptions in the attenuation correction. Colors are the same as the top panel.
    Bottom: The ratios of attenuation corrected spectra to their corresponding true power spectra. The black dashed line is for the mean of our 1000 light cones while the darker and lighter gray regions show the range containing 68\% and 95\% of light cones, illustrating the uncertainty introduced by attenuation corrections to an individual light cone. The ratios for 10 individual light cones are shown by colored lines.}
    \label{fig:attenmodel}
\end{figure}

\begin{deluxetable}{cccc}
\tablecaption{Summary of simulation parameters used in  Section~\ref{ss:systematics}\label{tab:models}. Attenuation corrections, when applied, assume a FWHM of 300 km s$^{-1}$ and $\sigma_z/(1+z)$ of 0.001, even when the input values of these parameters differ.}
\tablehead{
    \colhead{Name} & \colhead{FWHM} & \colhead{$\frac{\sigma z}{(1+z)}$} & \colhead{Corrected} }
\startdata
    True & 0 & 0 & No \\
    Attenuated & 300 & 0.001 & No \\
    Corrected & 300 & 0.001 & Yes \\
    \hline
    \multirow{2}{*}{$\sigma_\mathrm{lw}$ range} & 100 & \multirow{2}{*}{0.001} & \multirow{2}{*}{Yes} \\
    & 500 & & \\
    \hline
    \multirow{2}{*}{$\sigma_z$ range} & \multirow{2}{*}{300} & 0.0007 & \multirow{2}{*}{Yes} \\
    & & 0.0013 &  \\
\enddata
\end{deluxetable}

As discussed in Section~\ref{ss:atten}, a number of factors are known to distort observed intensity mapping power spectra from their intrinsic shape and amplitude. Here we explore this in greater depth by implementing the following effects in our model:
\begin{enumerate}
    \item Redshift uncertainty: we account for the decorrelation caused by redshift uncertainties by assigning each simulated galaxy an observed redshift drawn from a Gaussian distribution with a width $\sigma_z/(1+z)$ and centered at the true redshift.
    \item CO line width: we simulate the spreading of CO signal across multiple spectral channels by assigning all galaxies a Gaussian line profile characterized by a single FWHM.
    \item Imperfect attenuation corrections: our attenuation corrections assume a redshift error and line width. If these assumptions are incorrect, they will can result in over- or under-correction of the attenuation. To account for this effect we run versions of our model where we increase or decrease the redshift errors and line widths, but still correct them assuming the values of $\sigma_z$ and $\sigma_{\rm lw}$ given in Section~\ref{ss:atten}.
\end{enumerate}
Our inclusion of these effects allows us to assess how well our attenuation corrections work and estimate the uncertainty introduced by this step in our analysis. 

In the upper panel of Figure~\ref{fig:attenmodel} we show the cross-spectra of a series of models run with different combinations of these effects. Table~\ref{tab:models} summarizes the parameters of each model.

The solid black line shows the ``true'' power spectrum, while the gray dashed line shows the power spectrum attenuated by 300 km s$^{-1}$ line widths and redshift errors of $\sigma_z/(1+z)=0.001$. Comparing these models illustrates the expected signal attenuation in absence of corrections. The attenuation increases with $k$, as our survey accesses low $k$ modes primarily perpendicular to the line of sight where the $k_z$ component is small and attenuation is minimal. However at large $k$ the power spectrum is attenuated by roughly 70\%. The dashed black line shows the power spectrum with our attenuation correction applied, demonstrating that we successfully recover most of the attenuated power.

The filled regions in the top and middle panels of Figure~\ref{fig:attenmodel} show the effects of choosing the wrong values of $\sigma_z$ or $\sigma_\mathrm{lw}$ when applying the attenuation correction. These simulations are run with line widths of 100 and 500 km s$^{-1}$ (yellow regions) or redshift uncertainties of $\sigma_z/(1+z)=0.0007$ or $0.0013$ (blue regions), but are attenuation corrected assuming the fiducial values from Section~\ref{ss:atten} (${\rm FWHM}=300$ km s$^{-1}$, $\sigma_z/(1+z)=0.001$). The middle panel shows that both of these effects result in factional errors of less than 20\%.

Finally, the bottom panel of Figure~\ref{fig:attenmodel} shows the ratio of the corrected power spectrum with the true power spectrum for 10 individual light cones, along with the distribution from all 1000 simulated fields. Averaged over all light cones, the correction works within a few percent, while the attenuation-corrected cross spectra for 68\% of individual light cones match the true power spectra to within $\pm25\%$ at a given $k$.

\subsubsection{Error Budget for SZA Cross Correlation}
\begin{figure}
    \centering
    \includegraphics[width=.45\textwidth]{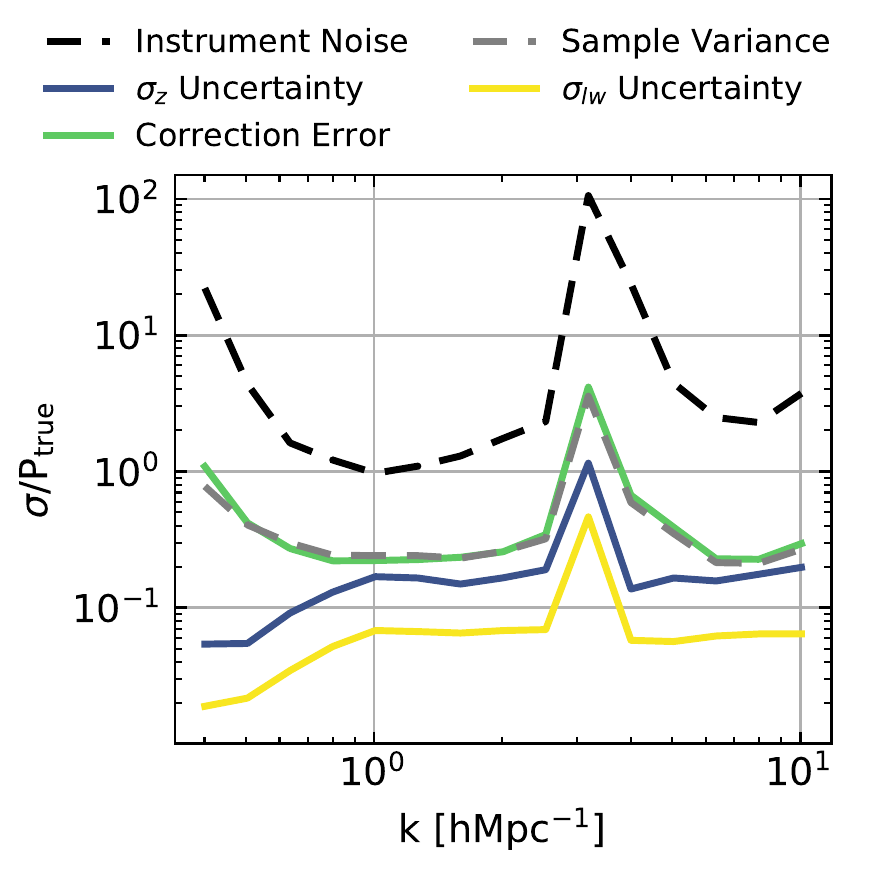}
    \caption{The fractional uncertainty in our power spectrum due to thermal noise (black dashed line), sample variance (gray), attenuation correction (green), and imperfect knowledge of galaxy redshift errors (blue) and linewidths (yellow).}
    \label{fig:szamodel}
\end{figure}

To this point, we have assumed uniform coverage of the uv-plane by our observations. In practice however, our data can only be used to measure the cross-spectrum at points in uv-space sampled by the SZA. We therefore recomputed our simulated power spectra using the same uv-sampling and cell-weighting as our data. 

Figure~\ref{fig:szamodel} shows the fractional uncertainty due to the attenuation correction, imperfect knowledge of $\sigma_z$ and $\sigma_{\rm lw}$, thermal noise and sample variance for our simulations including the SZA uv-coverage. The $\sigma_z$ uncertainties correspond to a $\pm30\%$ change $\sigma_z$, while the $\sigma_{\rm lw}$ uncertainties correspond to $\pm200$ km s$^{-1}$ in FWHM. The sample variance is computed from our ensemble of simulated light cones.

At all $k$, the thermal noise is dominant, and equals or exceeds the expected signal. Sample variance and discrepancies between the true and corrected power spectrum are the largest of the remaining sources of uncertainty at around 20-30\% for most $k$, while the remaining effects are smaller. The sample variance and attenuation correction-related uncertainties and SZA residuals all also scale with the expected signal, and as our fiducial model is at the bright end of the range of possible models (see Section~\ref{ss:modcomparisons}), these errors would produce an even smaller contribution to the total error budget for fainter signals. We therefore consider only thermal noise in the analysis of our observed power spectrum, but note that these other errors will be important for more sensitive future experiments.

\section{Discussion} \label{sec:discussion}

Our cross-power spectrum includes contributions from the shot and clustering components of the power spectrum. At small $k$, 
the clustering term is expected to exceed the shot power, and the difference in the $k$-space shape of these contributions makes them potentially separable. We therefore construct a two-component model for our observed spectrum. We compute the linear matter power spectrum appropriate for each window in our data using the \textsc{halomod} package \citep{halomod} and combine these into a single spectrum using the same weights as our data. We then fit two parameters: $\langle T_\mathrm{CO}\rangle b_\mathrm{CO}b_\mathrm{gal}$ (the factor scaling $P_\mathrm{lin}$) and $P_\mathrm{shot,\times}$ (a constant contribution at all $k$). 

\begin{figure}
    \centering
    \includegraphics[width=.45\textwidth]{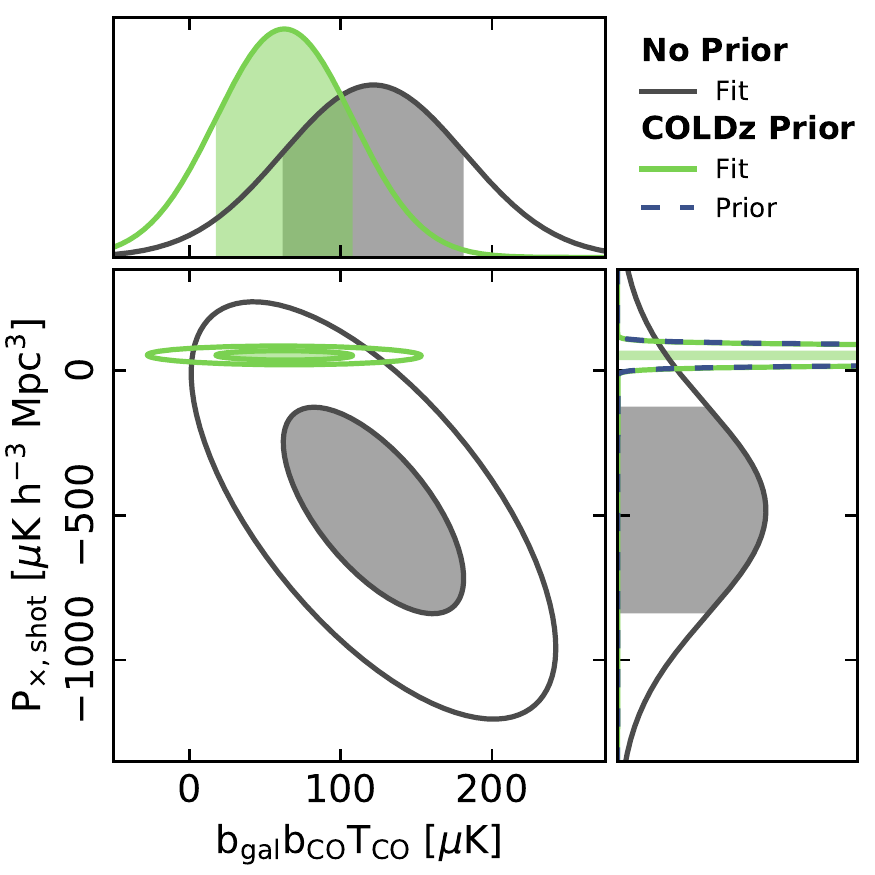}
    \caption{Lower left: the joint distribution of $\langle T_\mathrm{CO}\rangle b_\mathrm{CO}b_\mathrm{gal}$ and $P_\mathrm{shot,\times}$ for our fits with no priors (gray) and with a prior on $P_\mathrm{shot,\times}$ based on the \citet{pavesi+18} stack (green). Contours show 1$\sigma$ and 2$\sigma$ levels of the distribution. 
    Upper left: the likelihood distribution of $b_\mathrm{gal}b_\mathrm{CO}\langle T_\mathrm{CO}\rangle$. The filled regions show the central 68\% of the probability distribution. Lower right: The distribution of $P_\mathrm{shot,\times}$. The blue dashed line shows the prior based on the COLDz stack.}
    \label{fig:fit}
\end{figure}

The gray curves in Figure~\ref{fig:fit} show the parameter distribution for this fitting procedure. We find \bbTfitnp{} and \pxfitnp{} (medians and limits containing the central 68\% of the marginalized probability distribution for each parameter). The fit is degenerate, and the data marginally favor a negative shot power, pushing the clustering term towards more positive values. Negative shot power would correspond to negative average CO luminosity, and is therefore nonphysical.

We can improve our fit by using an external constraint on the average line luminosity of galaxies in our catalog to construct a prior on the $P_\mathrm{shot,\times}$. \citet{pavesi+18} performed a stacking analysis of CO(1-0) emission from GOODS-N galaxies at $2.0<z<2.8$ using a catalog similar to ours. Their stack was constructed using only objects undetected in their CO data, so we correct the average luminosity to include the two detected galaxies in their sample and use this to place a prior of $P_\mathrm{shot,\times}=54\pm16$ \xunits{}. Rerunning the fit then gives \bbTfitcoldz{}. The full parameter distribution is shown in green in Figure~\ref{fig:fit}. We find good agreement between our results with the COLDz prior and a prior that simply requires a positive shot power term.

\subsection{Comparison to Literature Models}\label{ss:modcomparisons}

\begin{figure}
    \centering
    \includegraphics[width=.45\textwidth]{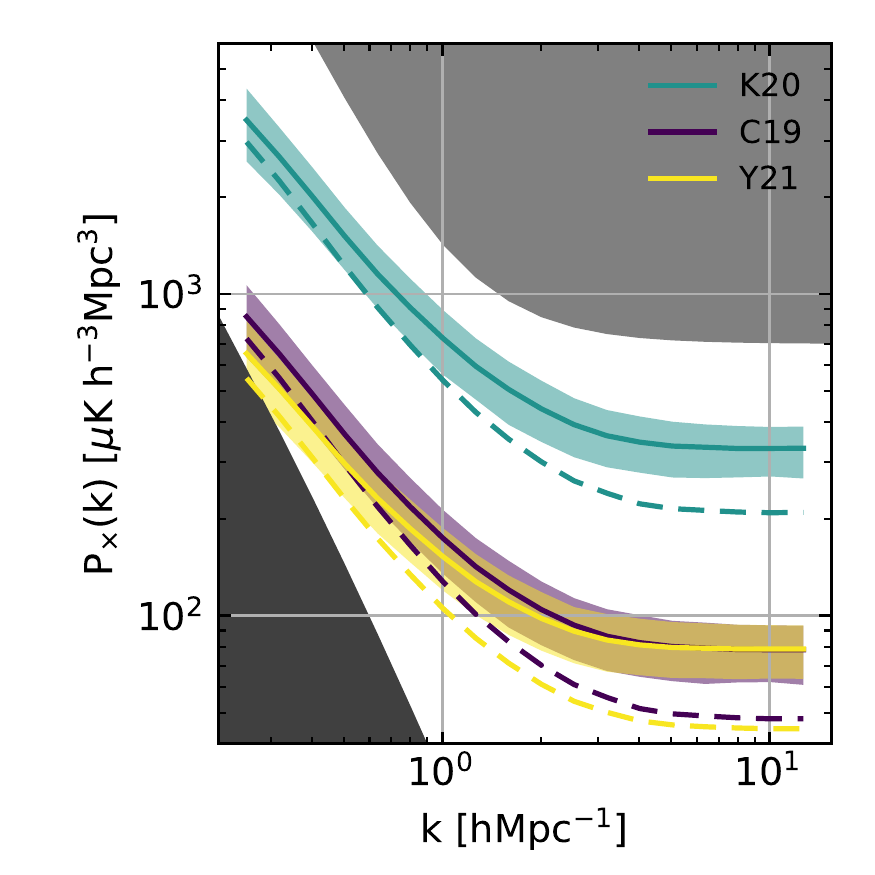}
    \caption{Cross-power spectra simulated using light cones from the TNG300 simulation and the \citetalias{keating+20} (blue-green), \citet[][purple]{chung+21_arxiv}, and \citet[][yellow]{yang+21b_arxiv} prescriptions for CO luminosity. Solid lines show the mean power spectrum of 1000 light cones and bands show the area containing 68\% of the realizations, while dashed lines show the result of using a galaxy catalog consisting of lower mass galaxies. The darker gray region at the lower left is ruled out by lower limits on $\langle T_\mathrm{CO}\rangle$ from \citet{decarli+20}. The lighter gray filled region at the upper right shows the parameter space excluded by our non-detection. }
    \label{fig:crossmodel}
\end{figure}

Models of the CO power spectrum disagree by as much as an order of magnitude in signal amplitude. We therefore re-run the mock observations described in Section~\ref{sec:model} with two additional models proposed in recent theoretical literature.

\citet{chung19b} use a double power law relating $M_{\rm halo}$ to $L_{\rm CO}$. Their fiducial parameters are chosen to approximately reproduce the $M_{\rm halo}$-$L_{\rm CO}$ relation of \citet{li+16} which in turn is calibrated using the halo mass to SFR relation of \citet{behroozi+13} and the SFR-$L_{\rm CO}$ correlation summarized in \citet{kennicutt98} and \citet{carilli+13}. This model includes a log-normal scatter of $\sigma_{\rm CO}=0.40$ similar to that in our fiducial model. As the results of the \citet{li+16} and \citet{chung19b} models are by design very similar, we only consider the more recent \citet{chung19b} model here.

We also consider the model of \citet{yang+21b_arxiv} who provide parametric formulas for $L_{\rm CO}$ as a function of halo mass and redshift calibrated to reproduce the results of the semi-analytic model (SAM) of far-infrared emission presented in \citet{yang+21a}.

Finally, in order to explore how our galaxy catalog affects the signal, we consider the results of modifying our optical galaxy catalog to contain fewer of the highest mass galaxies. We compute the cross power spectra of all three CO models with this modified galaxy catalog along with the fiducial catalog described in Section~\ref{ss:moddescription}.

The simulated cross-spectra are shown in Figure~\ref{fig:crossmodel}. We use our 2$\sigma$ sensitivity levels for the shot and clustering terms to set upper limits on the signal amplitude. We also use the lower limit on $\rho_\mathrm{mol}$ at $z\sim2.5$ reported by \citet{decarli+20} and our estimates of $b_{\rm gal}$ and $b_{\rm CO}$ from Section~\ref{ss:Trho} to estimate a lower limit on the clustering power. The regions excluded by these limits are shown in gray in Figure~\ref{fig:crossmodel}. 

The models based on \citet{chung19b} and \citet{yang+21b_arxiv} are well below the sensitivity of the current data set. The SAM underlying the \citet{yang+21b_arxiv} model was compared to recent CO intensity mapping observations in \citet{breysse+21_arxiv} and found to be in tension with current observational results. The \citet{chung19b} model is also in moderate tension with the tentative detection of the CO auto-power in COPSS, producing an auto-power signal approximately an order of magnitude fainter. Our simulations show that in cross-correlation both models also run up against the lower limit on the clustering power from direct detection. These discrepancies suggest that the power spectra of these two models represent lower limits on the range of possible signals. 

On the other hand, the \citetalias{keating+20} model lies near the upper edge of the allowed parameter space. Averaging across the full range in $k$, the expected SNR on $P_\mathrm{tot,\times}$ for \citetalias{keating+20} and our fiducial galaxy catalog ranges from $1.6$ to $2.4$ across the central 68\% of light cones. For the lower mass catalog this range is $1.1$ to $1.7$.

It is noteworthy that the COPSS auto-power measurement on which the \citetalias{keating+20} model is based consisted of ten times more CO data than the subset used here and achieved only a SNR of 2.3 for the auto-power, while the expected cross-power SNR for the same model with our smaller data set is 2.0. If we could use the full COPSS survey area in cross-correlation, or if all survey time had been dedicated to the GOODS-N field, we would expect to detect the cross-power for this model at ${\rm SNR}>6$ and be able to confidently fit the shot and clustering terms. While no re-imagining of COPSS would detect the \citet{chung19b} and \citet{yang+21b_arxiv} models, a number of ongoing CO intensity mapping experiments studying multiple transitions of the CO line will achieve much greater depths than the data presented here. Many of these studies are targeting fields with significant spectroscopic redshift information \citep{chung+19a,sun+21}. These experiments can therefore be expected to detect CO-galaxy cross-power and, for optimistic forecasts, constrain its shape in order to extract valuable physical information about the high redshift galaxy population. It is possible that such results could be rendered well before these experiments reach the sensitivity required to measure the auto-power spectrum.

\subsection{Limits on Average CO Luminosities}\label{ss:Lco}

Using equation~\ref{eq:pxshot} we can convert the shot component of the cross-power into an average CO luminosity of the galaxy catalog used for cross-correlation. Assuming the shot power dominates the power spectrum across the full range of $k$ probed here, our upper limit on the total power corresponds to a 2$\sigma$ upper limit on the mean CO luminosity of our galaxies of \lumlimittp{}. Our best-fit power spectrum (with no prior on $P_\mathrm{shot,\times}$) allows us to separate the shot and clustering terms and results in an upper limit of \lumlimitfit{}. These are comparable to the limit from our stack of \lumlimstack{}.

\begin{figure}
    \centering
    \includegraphics[width=.45\textwidth]{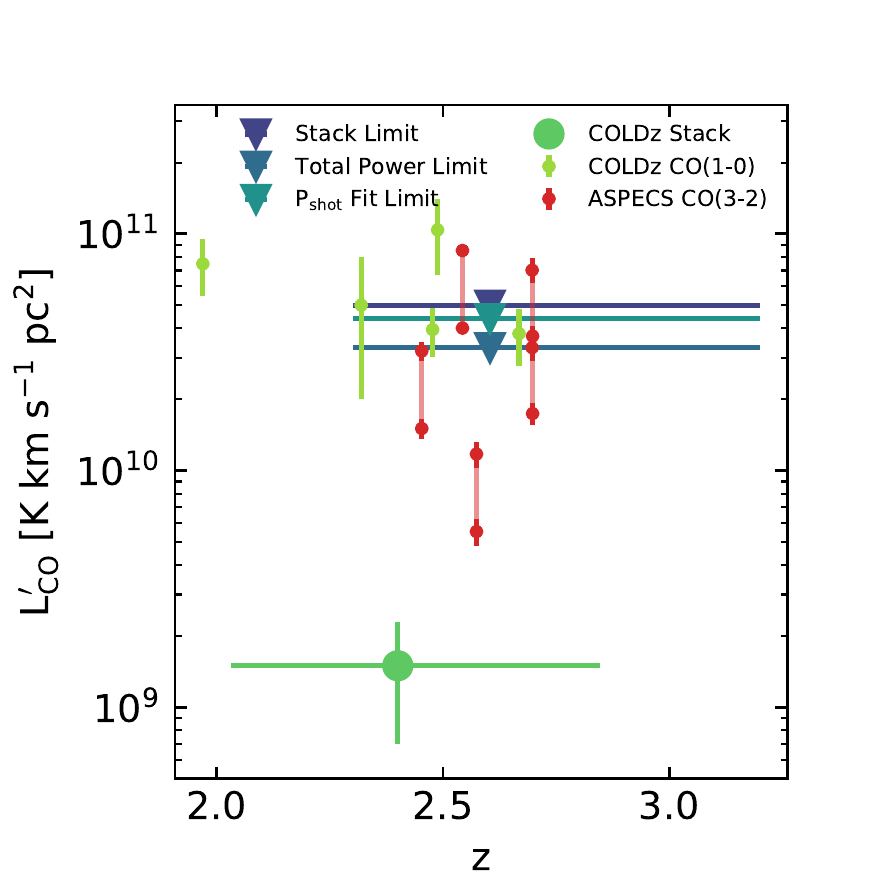}
    \caption{Upper limits on the CO luminosity of galaxies in our catalog derived from both our power spectrum analysis and our stacked spectrum (blue), compared to literature luminosity measurements at comparable redshifts. The stack of 78 GOODS-N galaxies from COLDz is shown as the large green point, while individually detected galaxies from the COLDz survey are shown in light green. Galaxies detected in CO(3-2) by the ASPECS survey are shown in red. As a range of values for the CO(3-2)/CO(1-0) line ratio have been reported, we convert each ASPECS luminosity to CO(1-0) assuming line ratios of 0.85 and 0.40 and show these two points connected by a vertical line for each source.}
    \label{fig:Lvsz}
\end{figure}

In Figure~\ref{fig:Lvsz} we present these upper limits on the mean CO luminosity of our catalog, along with the CO luminosities of confirmed galaxies detected in blind searches for CO emission at comparable redshifts. \citet{pavesi+18} reported CO(1-0) emission from five galaxies at $2.0<z<2.8$ spread over regions GOODS-N and COSMOS surveyed with the VLA. \citet{gonzalez-lopez+19} reported an additional five galaxies in CO(3-2) at $2.0<z<3.1$ detected in the Hubble Ultra Deep Field with ALMA. We convert CO(3-2) luminosities to CO(1-0) assuming a CO(3-2) to CO(1-0) luminosity ratio of 0.85 or 0.40 \citep{daddi+15,boogaard+20}. Our upper limits sit around the median luminosity of these direct detections. These results are in agreement with \citet{pavesi+18}, who performed a stacking analysis in GOODS-N using a similar galaxy catalog \citep{skelton+14}, and measured a mean luminosity about an order of magnitude below our limit.

\subsection{Interpreting the Clustering Constraint}\label{ss:rhoH2}

Measuring the average abundance of molecular gas in galaxies too faint for direct study at high redshift is a major objective of line intensity mapping. Assuming a constant molecular gas mass to CO luminosity ratio $\alpha_\mathrm{CO}$, the molecular gas density can be written in terms of the mean CO brightness temperature as
\begin{equation}\label{eq:rhoH2}
    \rho_\mathrm{mol} = \alpha_\mathrm{CO}\frac{H(z)}{(1+z)^2}\langle T_\mathrm{CO}\rangle
\end{equation}
where $H(z)$ is the Hubble parameter at redshift $z$. 

Intensity mapping measurements constrain $\langle T_\mathrm{CO}\rangle$ through the clustering term of the power spectrum which, for the cross-spectrum is proportional to $b_\mathrm{gal} b_\mathrm{CO} \langle T_\mathrm{CO}\rangle$. If $b_\mathrm{gal}$ and $b_\mathrm{CO}$ are known or can be measured, then a constraint on the amplitude of the clustering power becomes a constraint on $\langle T_\mathrm{CO}\rangle$ alone. 

In the remainder of this Section, we discuss our procedure for determining $b_\mathrm{gal}$ and $b_\mathrm{CO}$ in order to constrain $\langle T_\mathrm{CO}\rangle$ from our cross-spectrum results alone and in combination with the COPSS auto-power spectrum.

\subsubsection{Galaxy Bias}\label{ss:bgal}

We can measure $b_{\rm gal}$ directly from our catalog by measuring the galaxy-galaxy auto power spectrum
\begin{equation}\label{eq:pgal}
    P_{\rm gal}(k,z) = b_{\rm gal}(z)^2 P_{\rm lin}(k,z) + P_{\rm shot,gal} ,
\end{equation}
where $P_{\rm shot,gal}$ is the galaxy shot power and, for individual spectral windows will be equal to $\overline{n}^{-2}$, the inverse of the galaxy number density squared. We compute the auto-power spectrum of our galaxy density grid using the estimator
\begin{equation}
    \mathcal{P}_{\rm gal}(\mathbf{k}) = \frac{1}{V_{\rm cat}}(\tilde{\delta}_N^*(\mathbf{k})\tilde{\delta}_N(\mathbf{k}))
\end{equation}
\begin{equation}
    P_{\rm gal}(k)=\langle \mathcal{P}_{\rm gal}(\mathbf{k}) \rangle_{\mathbf{k\cdot k}=k^2}
\end{equation}
and fit the one dimensional galaxy power spectrum with clustering and shot components. The resulting estimate of $b_{\rm gal}$ is $3.5^{+0.6}_{-0.8}$.

The bias of OIR selected, star forming galaxies has been studied extensively, allowing us to cross-check our fit with literature results. \citet{durkalec+15} measure the galaxy bias of a large sample of rest frame ultra-violet selected galaxies with spectroscopic redshift from the VIMOS Ultra Deep Survey (VUDS), and find $b_\mathrm{gal}\sim2.7$ to $2.8$ (depending on fitting method) at a mean redshift of $z=2.95$. \citet{alonso+21_arxiv} measure $b_\mathrm{gal}\sim2.8$ to $3.0$ for Ly$\alpha$ emitting galaxies at a median redshift $z=3.8$. \citet{geach+12} find $b_\mathrm{gal}\sim2.4$ for H$\alpha$ emitting galaxies at $z=2.2$. \citet{adelberger+05} find $b_\mathrm{gal}\sim2.1$, $2.4$, and $2.6$ at $z=1.7$, $2.2$, and $2.9$ for galaxies identified using $U_nG\mathcal{R}$ color selection. \citet{kashikawa+06} find $b_\mathrm{gal}\sim5$ for $z=4.1$ luminous blue galaxies (LBGs).

These results are generally consistent with our own fit. The overall convergence of these results around $b_\mathrm{gal}\sim3$ masks a number of difficulties in determining $b_\mathrm{gal}$ for a given galaxy sample. The galaxy bias is known to be a function of galaxy properties, with brighter, more massive, and redder galaxy samples all tending to be more clustered \citep{kashikawa+06,khostovan+19}. The bias also appears to increase from redshifts $\sim2$ to $6$ \citep{durkalec+15,khostovan+19}. In particular, because of the inhomogeneous nature of the galaxy catalog we assemble, the appropriate galaxy bias for our analysis may differ from these results. In addition, galaxy clustering studies typically use correlation function (i.e. real space) analyses to determine $b_\mathrm{gal}$ which may introduce different systematics compared to our power spectrum derived constraint. Therefore we use the value derived directly from our catalog in the following analysis.

\subsubsection{CO Tracer Bias}\label{ss:bco}

The CO tracer bias is less studied. In theory it can be extracted from anisotropies in the three dimensional CO power spectrum, however such a measurement requires much more sensitive data than is currently available. 

Instead we can compute $b_{\rm CO}$ as
\begin{equation}
    b_\mathrm{CO} = \frac{\int L_\mathrm{CO}(M) b(M) \frac{dn}{dM} dM}{\int L_\mathrm{CO}(M) \frac{dn}{dM} dM}
\end{equation}
where $b(M)$ is the mass dependent halo bias, $dn/dM$ is the halo mass function, and $L_\mathrm{CO}(M)$ is the halo mass to CO luminosity relation. To estimate $b_\mathrm{CO}$ we use the halo mass function of \citet{tinker+08,tinker+10}, the halo bias prescription of \citet{tinker+10}, and our halo mass to CO luminosity model outlined in Section~\ref{sec:model} and Equation~\ref{eq:k20model}. This gives $b_\mathrm{CO}=2.3$. Using a similar procedure, but a different halo mass to CO luminosity scaling \citet{chung+19a} find $b_\mathrm{CO}=2.7$. We take the mean of these two values as our estimate.

\subsubsection{Joint Fit of Galaxy, CO, and Cross Power Spectra}

The CO auto spectrum and galaxy-CO cross spectrum both depend on $\langle T_{\rm CO}\rangle$, but with different exponents and different combinations of bias terms: $P_{\rm cluster,CO}\propto \langle T_{\rm CO}\rangle^2 b_{\rm CO}^2$ and $P_{\rm cluster,\times}\propto \langle T_{\rm CO}\rangle b_{\rm CO} b_{\rm gal}$. Meanwhile the galaxy auto spectrum contains additional information about the galaxy bias term with $P_{\rm cluster,gal}\propto b_{\rm gal}^2$. Fitting all three spectra jointly can extract the maximum information from an intensity mapping data set. 

We therefore perform a joint fit to the final COPSS CO auto spectrum \citepalias{keating+20}, our cross spectrum, and our galaxy auto spectrum using a model with five parameters: $b_{\rm CO}\langle T_{\rm CO}\rangle$, $b_{\rm gal}$, $P_{\rm shot,CO}$, $P_{\rm shot,\times}$, and $P_{\rm shot,gal}$. To account for sample variance between the COPSS spectrum measured over many fields, we add a nuisance term in our model which allows $b_{\rm CO}\langle T_{\rm CO}\rangle$ to vary between the CO auto spectrum and the cross spectrum. For this term we set a Gaussian prior with a mean difference of zero and a standard deviation of 10\% of $b_{\rm CO}\langle T_{\rm CO}\rangle$, chosen based on the sample variance model of \citet{keenan+20}. We fit this model using a Markov Chain Monte Carlo procedure implemented with the {\sc emcee} package \citep{foreman-mackey+13}.

\begin{figure}
    \centering
    \includegraphics[width=.45\textwidth]{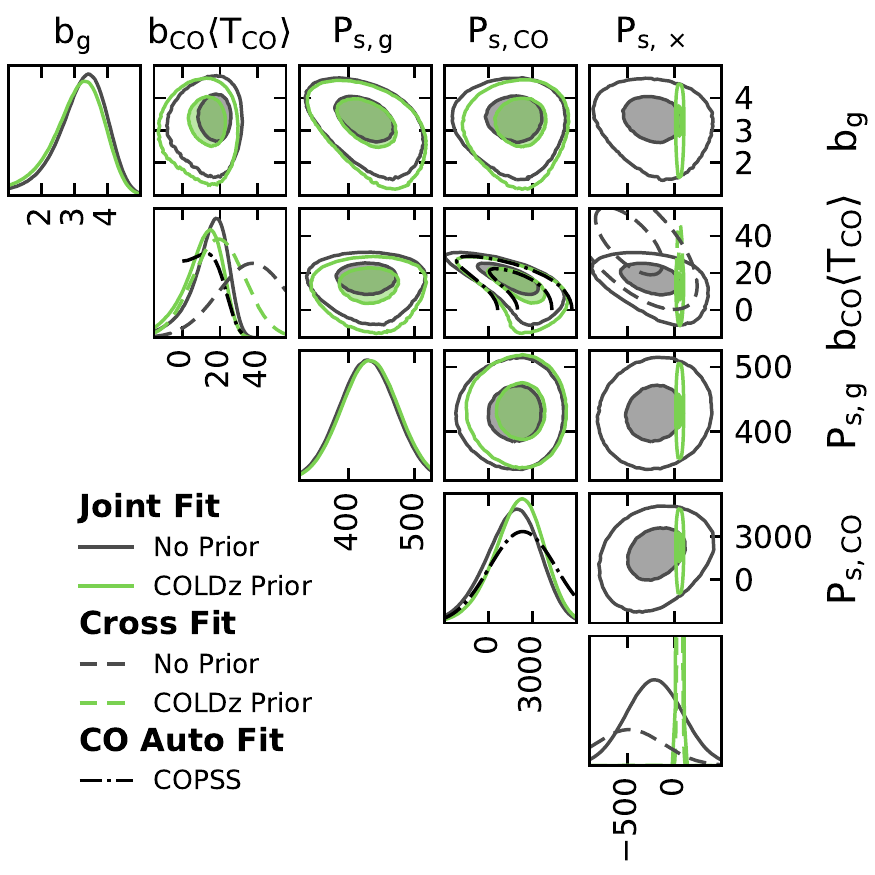}
    \caption{The first panels of each row show the marginalized probability distributions for $b_{\rm gal}$, $b_{\rm CO}\langle T_{\rm CO}\rangle$, $P_{\rm shot,gal}$, $P_{\rm shot,CO}$, and $P_{\rm shot,\times}$ (from top to bottom) based on our joint fitting of the CO, galaxy, and cross power spectra. The remaining panels show the joint distributions of pairs of parameters, with contours at the 1$\sigma$ and 2$\sigma$ levels of the distribution. Gray shows the fit performed with no prior on the cross-shot power, while green shows fits using the \citet{pavesi+18} stack as a prior. Dashed lines in the $b_{\rm CO}\langle T_{\rm CO}\rangle$ and $P_{\rm shot,\times}$ show the results from fitting only the cross-power spectrum highlighting the improved constraining power of the joint fit. Black dot-dashed lines in the $b_{\rm CO}\langle T_{\rm CO}\rangle$, and $P_{\rm shot,CO}$ show the results of fitting only the COPSS CO auto-power spectrum. The auto spectrum alone only constrains $b_{\rm CO}^2\langle T_{\rm CO}\rangle^2$ and therefore the contours do not extend below $b_{\rm CO}\langle T_{\rm CO}\rangle=0$.}
    \label{fig:mcmcfit}
\end{figure}

The resulting five-parameter fit is shown and compared to fits to the cross spectrum or CO auto spectrum alone for parameters that can be constrained by individual spectra in Figure~\ref{fig:mcmcfit}. The results are largely consistent with the individual spectrum fits presented at the beginning of this section and in \citet{keating+20}. However, the joint fit results in significantly tighter constraints on $b_{\rm CO}\langle T_{\rm CO}\rangle$ and rules out extremely negative values of $P_{\rm shot,\times}$ even without the use of a prior on this term. We summarize the fit results in Table~\ref{tab:fits}.

\begin{deluxetable}{lccc}
\tablecaption{Estimates for power spectrum terms based on fits to $P_{\rm CO}$, $P_{\rm \times}$, and/or $P_{\rm gal}$. Values are shown both with and without use of a prior on $P_{\rm shot,\times}$ of $54\pm16$ \xunits{} derived from the results of \citet{pavesi+18}. \label{tab:fits}}
\tablehead{
    \colhead{Parameter} & \colhead{Units} & \multicolumn{2}{c}{Value} \\
    \colhead{ } & \colhead{ } & \colhead{(w/ prior)} & \colhead{(no prior)}
}
\startdata
    & & \multicolumn{2}{c}{Joint fit of $P_{\rm CO}$, $P_{\rm \times}$, $P_{\rm gal}$} \\
    \hline
    $b_{\rm gal}$ & & $3.1^{+0.7}_{-0.9}$ &  $3.3^{+0.7}_{-0.9}$ \\
    $b_{\rm CO} \langle T_{\rm CO}\rangle$ & $\mu$K & $12^{+9}_{-12}$ & $15^{+8}_{-11}$ \\
    $P_{\rm shot,gal}$ & h$^{-3}$Mpc$^3$ & $430\pm40$ & $430\pm40$ \\
    $P_{\rm shot,CO}$ & \aunits{} & $2100^{+1600}_{-1800}$ &  $1700^{+1700}_{-1900}$ \\
    $P_{\rm shot,\times}$ & \xunits{} & $53\pm16$ & $-210\pm300$  \\
    \hline
    & & \multicolumn{2}{c}{Fit of $P_{\rm \times}$} \\
    \hline
    $b_{\rm gal} b_{\rm CO} \langle T_{\rm CO}\rangle$ & $\mu$K & \bbTfitcoldznum{} & \bbTfitnpnum{} \\
    $P_{\rm shot,\times}$ & \xunits & $53\pm16$ & \pxfitnpnum{} \\
    \hline
    & & \multicolumn{2}{c}{Fit of $P_{\rm CO}$} \\
    \hline
    $b_{\rm CO} \langle T_{\rm CO}\rangle$ & $\mu$K & --- & $<31$ \\
    $P_{\rm shot,CO}$ & \aunits{} & --- & $2000^{+1100}_{-1200}$ \\
    \hline
    & & \multicolumn{2}{c}{Fit of $P_{\rm gal}$} \\
    \hline
    $b_{\rm gal}$ & &  --- & $3.5^{+0.6}_{-0.8}$ \\
    $P_{\rm shot,gal}$ & h$^{-3}$Mpc$^3$  & --- & $430\pm40$ \\
    \hline
\enddata
\end{deluxetable}

\subsubsection{Limits on the Mean Molecular Gas Abundance}\label{ss:Trho}

\begin{figure}
    \centering
    \includegraphics[width=.45\textwidth]{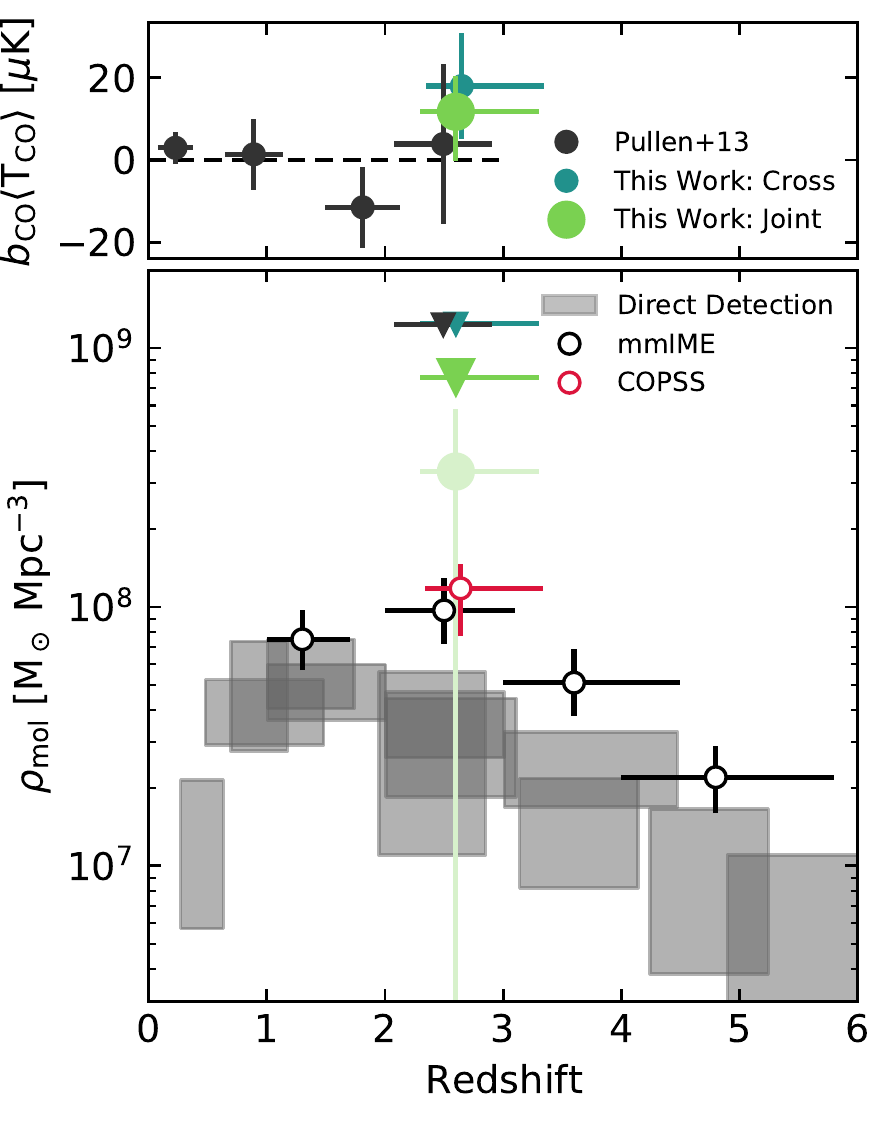}
    \caption{Measurements of the CO mean brightness temperature and molecular gas density presented as a function of redshift based on various CO survey methodologies. Top: $b_\mathrm{CO}\langle T_\mathrm{CO}\rangle$ constraints from CO-quasar cross correlation measurements by \citet[][black]{pullen+13} and the results reported here based on CO-galaxy cross correlation (blue-green point) and based on a joint analysis of the CO and galaxy auto and cross power spectra (large green point).
    Bottom: Our upper limits on $\rho_\mathrm{mol}$ (colored triangles) along with the upper limit of \citet[][black triangle]{pullen+13} and lower limits from direct detection in CO deep fields \citep[gray boxes;][]{decarli+20,lenkic+20,riechers+18}. Open circles correspond to CO auto-power spectrum experiments COPSS (black) and mmIME (red), which are used to fit a halo-mass to CO luminosity relation and then integrated over the halo mass function to extract $\langle T_{\rm CO}\rangle$. We also show the central value and 1$\sigma$ uncertainties of our measurement as the light green point but note that this point is not a significant detection.}
    \label{fig:rhoh2}
\end{figure}

Based on our fit to the cross spectrum alone, we derive a constraint on $\langle T_\mathrm{CO}\rangle$ of \Tvalnum{} or an upper limit of \Tlim{} (assuming the values of $b_{\rm gal}$ and $b_{\rm CO}$ from Sections~\ref{ss:bgal} and~\ref{ss:bco} and using the \citet{pavesi+18} prior on the cross shot power). Our joint fit to the cross and auto spectra allows us to derive a tighter constraint of  $\langle T_\mathrm{CO}\rangle=4.7^{+3.5}_{-4.8}$ $\mu$K and an upper limit of $\langle T_\mathrm{CO}\rangle<10.9$ $\mu$K (again using the \citet{pavesi+18} prior and assuming the $b_{\rm CO}$ from Section~\ref{ss:bco}, but now deriving $b_{\rm gal}$ directly from the fit). The inclusion of the auto power spectrum results in a factor of two reduction in our upper limit.

\citet{pullen+13} cross-correlated quasars with the cosmic microwave background maps of WMAP to constrain the CO-quasar cross power spectrum. In the upper panel of Figure~\ref{fig:rhoh2}, we show their reported limits on $b_\mathrm{CO}\langle T_\mathrm{CO}\rangle$, compared with our measurement. Our uncertainties are roughly a factor of two smaller than those reported by \citet{pullen+13}, largely due to our access to a three dimensional CO data set optimized for intensity mapping analyses, compared to the two dimensional maps available from WMAP.

To translate our results into mean molecular gas densities, we must assume a value of $\alpha_{\rm CO}$. Star forming galaxies at $z\gtrsim1$ have been found to have Milky Way-like $\alpha_{\rm CO}$ \citep{cassata+20,carleton+17,daddi+10}. This class of galaxy is expected to account for a significant fraction of the total CO luminosity  \citep{uzgil+19,inami+20}, and therefore we follow other recent works on the mean molecular gas density and adopt $\alpha_\mathrm{CO}\sim3.6$ M$_\odot$ (K km s$^{-1}$ pc$^2$)$^{-1}$ \citep{daddi+10}. For our cross spectrum-only constraint, this results in an upper limit of \rholim{} at $z\sim 2.6$. The corresponding limit for our joint fit is $\rho_{\rm mol}<7.7\times10^8$ M$_\odot$ Mpc$^{-3}$.

In the lower panel of Figure~\ref{fig:rhoh2} we compare our $\rho_\mathrm{mol}$ constraints to a number of literature results.

Multiple recent studies have produced deep spectroscopic (sub-)millimeter maps of large regions of sky in order to search individual CO emission lines \citep{pavesi+18,gonzalez-lopez+19}. The sources detected by these studies probe the contribution to $\langle T_\mathrm{CO}\rangle$ from the portion of the CO luminosity function above the survey detection threshold, and therefore represent lower limits on $\langle T_\mathrm{CO}\rangle$ and $\rho_\mathrm{mol}$ \citep{decarli+20,decarli+19,riechers+18,lenkic+20}. In Figure~\ref{fig:rhoh2} we plot these lower limits as gray boxes. Stacking on optically selected galaxies has also been used to explore the fraction of $\rho_\mathrm{mol}$ accounted for by direct detections in CO deep fields, suggesting that they account for around 50\% \citep{inami+20,walter+20}. However, these measurements are only sensitive to the galaxies included in the stacks, and galaxy catalogs are incomplete at $z\sim2$-$3$ where determining spectroscopic redshifts is difficult. 

\citetalias{keating+20} fit models of the halo mass to CO luminosity relation to the shot power measured in the mmIME and COPSS CO auto spectra, and combined these with the halo mass function to produce estimates of $\rho_\mathrm{mol}$. These results, shown as open circles in Figure~\ref{fig:rhoh2}, show a molecular density higher than reported by direct detection studies at $z\sim2.5$. The auto-power spectra of mmIME and COPSS contain information about all emitting galaxies, however the shot power is most sensitive to the brightest objects \citep{keenan+20}, and the constraints are sensitive to modeling assumptions \citep{breysse+21_arxiv}. 

On the other hand, our clustering power measurements are sensitive to all CO emission without any bias towards bright objects, and are also largely independent of modeling assumptions. This makes them the most reliable tracer of the contribution of faint galaxies and allows us to set a robust upper limit on the mean molecular gas density. While our survey is not particularly constraining at the current sensitivity, it confirms that the total molecular gas density cannot be much more than an order of magnitude above the direct detection limits. On the other hand, we cannot rule out the possibility that galaxies below the direct detection threshold represent a substantial contribution to the total molecular gas abundance, as suggested by the \citet{keating+20} auto-power results. Upcoming line intensity mapping studies will have much greater sensitivity than the results presented here, and should help to fully contextualize direct detection results.

\section{Conclusion}\label{sec:conclusion}

We have presented upper bounds on the CO-galaxy cross-power spectrum at $z\sim3$. We find a $2\sigma$ upper limit on the band-averaged cross-power of \pxlimit{}. This limit lies near models for the cross-power based on the tentative CO auto-power measurement of \citetalias{keating+16}. This suggests that the cross-power may be detectable in current intensity mapping data sets with modest increases in the coverage of spectroscopic redshift catalogs, and should easily be within the reach of the next generation of line intensity mapping experiments, which are currently collecting data. Our measurement is sensitive to the CO emission of {\it all} galaxies, allowing us to set a robust upper limit on the mean CO brightness temperature and constrain the mean molecular gas abundance at $z\sim2.6$ to be $\rho_{\rm mol}<7.7\times10^8$ M$_\odot$ Mpc$^{-3}$. This is a factor of two deeper than previous cross-correlation-based constraints.

Cross-correlation can be used to check for systematics and foregrounds in intensity mapping auto-power measurements. We demonstrate this by verifying the data reduction and cleaning of the CO Power Spectrum Survey auto-power measurement. In particular, we search for and find no evidence of bright CS or HCN foregrounds. We also demonstrate the effectiveness of using cross-correlation to remove instrumental effects, finding that strong ground-correlated emission in the COPSS data is cleaned to below the noise level by cross-correlation even with no other cleaning applied.

COPSS was a pathfinder intensity mapping experiment, conducted without the benefit of optimized instrumentation. The tentative detection of the CO auto-power spectrum in \citet{keating+16}, and the constraining limits on the CO-galaxy cross-power spectrum reported here demonstrate the power of the intensity mapping technique for studying galaxy evolution at high redshift, and in regimes not accessible via traditional observing modes. Since the completion of COPSS, a number of purpose built intensity mapping instruments have been fielded and are now collecting early science data. These instruments will greatly improve the sensitivity of intensity mapping analyses, resulting in competitive constraints on the CO luminosity of average galaxies, the evolution of the mean molecular gas density, and a wide array of astrophysics and cosmology.

\acknowledgments
RPK would like to thank P. Behroozi, X. Fan, R. Kennicutt, and E. Krause for guidance in the process of conducting this research, E. Mayer for useful feedback on the manuscript, and J. Keenan for support in concluding this project. The authors would like to thank the organizers of the UChicago/KICP Line Intensity Mapping Workshop, where discussions related to this work greatly improved the quality of the final paper. RPK was supported by the National Science Foundation through Graduate Research Fellowship grant DGE-1746060. DPM and RPK were supported by the National Science Foundation through CAREER grant AST-1653228.

\vspace{5mm}
\facilities{CARMA/SZA}

\appendix
\section{Additional Details on Ground-Contamination}\label{appendix:auto_contamination}

\begin{figure}
    \centering
    \includegraphics[width=.45\textwidth]{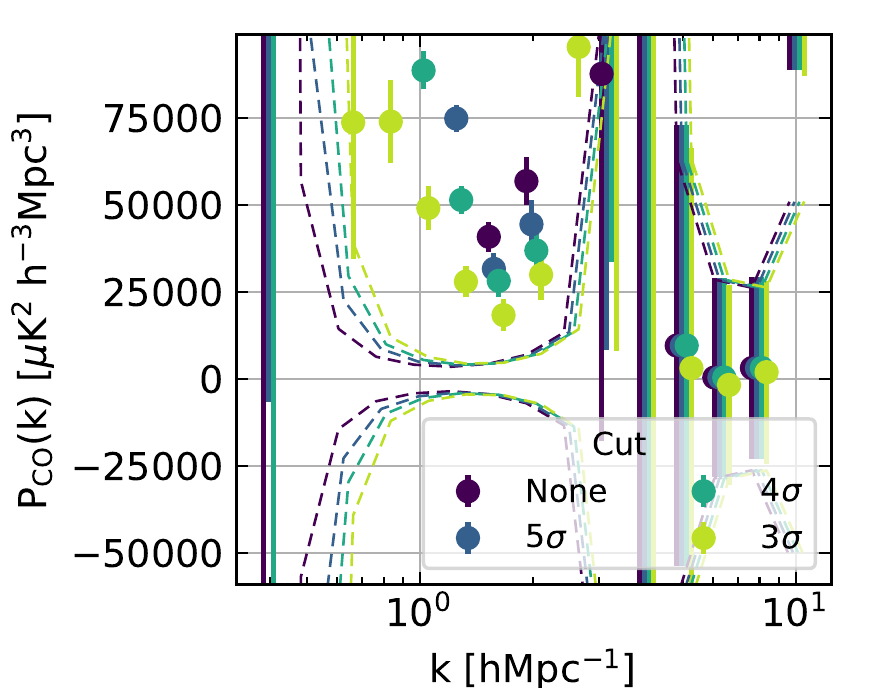}
    \caption{The auto-power spectrum for the GOODS-N field reproduced using a range of different significance cuts to clean the COPSS 30~GHz data. Values for each cut are offset slightly in k for clarity. Contamination is significant for all cuts and changes considerably depending on the level of cleaning performed.}
    \label{fig:autocuts}
\end{figure}

In Section~\ref{sec:validation} we demonstrated that the cross-spectrum is largely unaffected by the ground-correlated emission contaminating the SZA data. For comparison, we show auto-power spectrum of the GOODS-N field without corrections for this contamination in Figure~\ref{fig:autocuts}. These are produced using the same $\sigma$-cuts as Figure~\ref{fig:cuts}. 

Two things are noteworthy, first, at all cuts the power from ground contamination in low-k modes is significant, and second, removing modes significantly alters the level of contamination and causes large changes in the power spectrum. These changes are not observed in the cross-power spectrum, suggesting that the level of contamination does not affect the cross-power spectrum results, and confirming that cross-correlation has removed the contamination.

\bibliography{refs}

\begin{thebibliography}{}
\expandafter\ifx\csname natexlab\endcsname\relax\def\natexlab#1{#1}\fi
\providecommand{\url}[1]{\href{#1}{#1}}

\bibitem[{{Adelberger} {et~al.}(2005){Adelberger}, {Steidel}, {Pettini},
  {Shapley}, {Reddy}, \& {Erb}}]{adelberger+05}
{Adelberger}, K.~L., {Steidel}, C.~C., {Pettini}, M., {et~al.} 2005, \apj, 619,
  697

\bibitem[{{Barger} {et~al.}(2008){Barger}, {Cowie}, \& {Wang}}]{barger+08}
{Barger}, A.~J., {Cowie}, L.~L., \& {Wang}, W.~H. 2008, \apj, 689, 687

\bibitem[{{Behroozi} {et~al.}(2013){Behroozi}, {Wechsler}, \&
  {Conroy}}]{behroozi+13}
{Behroozi}, P.~S., {Wechsler}, R.~H., \& {Conroy}, C. 2013, \apj, 770, 57

\bibitem[{{Boogaard} {et~al.}(2020){Boogaard}, {van der Werf}, {Weiss},
  {Popping}, {Decarli}, {Walter}, {Aravena}, {Bouwens}, {Riechers},
  {Gonz{\'a}lez-L{\'o}pez}, {Smail}, {Carilli}, {Kaasinen}, {Daddi}, {Cox},
  {D{\'\i}az-Santos}, {Inami}, {Cortes}, \& {Wagg}}]{boogaard+20}
{Boogaard}, L.~A., {van der Werf}, P., {Weiss}, A., {et~al.} 2020, \apj, 902,
  109

\bibitem[{{Breysse} \& {Alexandroff}(2019)}]{breysse+19}
{Breysse}, P.~C., \& {Alexandroff}, R.~M. 2019, \mnras, 490, 260

\bibitem[{{Breysse} {et~al.}(2014){Breysse}, {Kovetz}, \&
  {Kamionkowski}}]{breysse+14}
{Breysse}, P.~C., {Kovetz}, E.~D., \& {Kamionkowski}, M. 2014, \mnras, 443,
  3506

\bibitem[{{Breysse} {et~al.}(2015){Breysse}, {Kovetz}, \&
  {Kamionkowski}}]{breysse+15}
---. 2015, \mnras, 452, 3408

\bibitem[{{Breysse} {et~al.}(2021){Breysse}, {Yang}, {Somerville}, {Pullen},
  {Popping}, \& {Maniyar}}]{breysse+21_arxiv}
{Breysse}, P.~C., {Yang}, S., {Somerville}, R.~S., {et~al.} 2021, arXiv
  e-prints, arXiv:2106.14904

\bibitem[{{Carilli} \& {Walter}(2013)}]{carilli+13}
{Carilli}, C.~L., \& {Walter}, F. 2013, \araa, 51, 105

\bibitem[{{Carleton} {et~al.}(2017){Carleton}, {Cooper}, {Bolatto}, {Bournaud},
  {Combes}, {Freundlich}, {Garcia-Burillo}, {Genzel}, {Neri}, {Tacconi}, {Sand
  strom}, {Weiner}, \& {Weiss}}]{carleton+17}
{Carleton}, T., {Cooper}, M.~C., {Bolatto}, A.~D., {et~al.} 2017, \mnras, 467,
  4886

\bibitem[{{Carucci} {et~al.}(2017){Carucci}, {Villaescusa-Navarro}, \&
  {Viel}}]{carucci+17}
{Carucci}, I.~P., {Villaescusa-Navarro}, F., \& {Viel}, M. 2017, \jcap, 2017,
  001

\bibitem[{{Cassata} {et~al.}(2020){Cassata}, {Liu}, {Groves}, {Schinnerer},
  {Ibar}, {Sargent}, {Karim}, {Talia}, {F{\`e}vre}, {Tasca}, {Lemaux},
  {Ribeiro}, {Fiore}, {Romano}, {Mancini}, {Morselli}, {Rodighiero},
  {Rodr{\'\i}guez-Mu{\~n}oz}, {Enia}, \& {Smolcic}}]{cassata+20}
{Cassata}, P., {Liu}, D., {Groves}, B., {et~al.} 2020, \apj, 891, 83

\bibitem[{{Chang} {et~al.}(2010){Chang}, {Pen}, {Bandura}, \&
  {Peterson}}]{chang+10}
{Chang}, T.-C., {Pen}, U.-L., {Bandura}, K., \& {Peterson}, J.~B. 2010, \nat,
  466, 463

\bibitem[{{Cheng} {et~al.}(2016){Cheng}, {Chang}, {Bock}, {Bradford}, \&
  {Cooray}}]{cheng+16}
{Cheng}, Y.-T., {Chang}, T.-C., {Bock}, J., {Bradford}, C.~M., \& {Cooray}, A.
  2016, \apj, 832, 165

\bibitem[{{Cheng} {et~al.}(2020){Cheng}, {Chang}, \& {Bock}}]{cheng+20}
{Cheng}, Y.-T., {Chang}, T.-C., \& {Bock}, J.~J. 2020, \apj, 901, 142

\bibitem[{{Chung}(2019)}]{chung19b}
{Chung}, D.~T. 2019, \apj, 881, 149

\bibitem[{{Chung} {et~al.}(2017){Chung}, {Li}, {Viero}, {Church}, \&
  {Wechsler}}]{chung+17}
{Chung}, D.~T., {Li}, T.~Y., {Viero}, M.~P., {Church}, S.~E., \& {Wechsler},
  R.~H. 2017, \apj, 846, 60

\bibitem[{{Chung} {et~al.}(2019){Chung}, {Viero}, {Church}, {Wechsler},
  {Alvarez}, {Bond}, {Breysse}, {Cleary}, {Eriksen}, {Foss}, {Gundersen},
  {Harper}, {Ihle}, {Keating}, {Murray}, {Padmanabhan}, {Stein}, {Wehus}, \&
  {COMAP Collaboration}}]{chung+19a}
{Chung}, D.~T., {Viero}, M.~P., {Church}, S.~E., {et~al.} 2019, \apj, 872, 186

\bibitem[{{Chung} {et~al.}(2021){Chung}, {Breysse}, {Tveit Ihle},
  {Padmanabhan}, {Silva}, {Bond}, {Borowska}, {Cleary}, {Eriksen}, {Foss}, {Ott
  Gundersen}, {Keating}, {Gahr Sturtzel Lunde}, {Stutzer}, {Viero}, {Watts}, \&
  {Kathrine Wehus}}]{chung+21_arxiv}
{Chung}, D.~T., {Breysse}, P.~C., {Tveit Ihle}, H., {et~al.} 2021, arXiv
  e-prints, arXiv:2104.11171

\bibitem[{{Cohn} {et~al.}(2016){Cohn}, {White}, {Chang}, {Holder},
  {Padmanabhan}, \& {Dor{\'e}}}]{cohn+16}
{Cohn}, J.~D., {White}, M., {Chang}, T.-C., {et~al.} 2016, \mnras, 457, 2068

\bibitem[{{Concerto Collaboration} {et~al.}(2020){Concerto Collaboration},
  {Ade}, {Aravena}, {Barria}, {Beelen}, {Benoit}, {B{\'e}thermin}, {Bounmy},
  {Bourrion}, {Bres}, {De Breuck}, {Calvo}, {Cao}, {Catalano}, {D{\'e}sert},
  {Dur{\'a}n}, {Fasano}, {Fenouillet}, {Garcia}, {Garde}, {Goupy}, {Groppi},
  {Hoarau}, {Lagache}, {Lambert}, {Leggeri}, {Levy-Bertrand},
  {Mac{\'\i}as-P{\'e}rez}, {Mani}, {Marpaud}, {Mauskopf}, {Monfardini},
  {Pisano}, {Ponthieu}, {Prieur}, {Roni}, {Roudier}, {Tourres}, \&
  {Tucker}}]{ade+20}
{Concerto Collaboration}, {Ade}, P., {Aravena}, M., {et~al.} 2020, \aap, 642,
  A60

\bibitem[{{Cunnington} {et~al.}(2019){Cunnington}, {Harrison}, {Pourtsidou}, \&
  {Bacon}}]{cunnington+19}
{Cunnington}, S., {Harrison}, I., {Pourtsidou}, A., \& {Bacon}, D. 2019,
  \mnras, 482, 3341

\bibitem[{{Daddi} {et~al.}(2010){Daddi}, {Bournaud}, {Walter}, {Dannerbauer},
  {Carilli}, {Dickinson}, {Elbaz}, {Morrison}, {Riechers}, {Onodera}, {Salmi},
  {Krips}, \& {Stern}}]{daddi+10}
{Daddi}, E., {Bournaud}, F., {Walter}, F., {et~al.} 2010, \apj, 713, 686

\bibitem[{{Daddi} {et~al.}(2015){Daddi}, {Dannerbauer}, {Liu}, {Aravena},
  {Bournaud}, {Walter}, {Riechers}, {Magdis}, {Sargent}, {B{\'e}thermin},
  {Carilli}, {Cibinel}, {Dickinson}, {Elbaz}, {Gao}, {Gobat}, {Hodge}, \&
  {Krips}}]{daddi+15}
{Daddi}, E., {Dannerbauer}, H., {Liu}, D., {et~al.} 2015, \aap, 577, A46

\bibitem[{{Decarli} {et~al.}(2016){Decarli}, {Walter}, {Aravena}, {Carilli},
  {Bouwens}, {da Cunha}, {Daddi}, {Ivison}, {Popping}, {Riechers}, {Smail},
  {Swinbank}, {Weiss}, {Anguita}, {Assef}, {Bauer}, {Bell}, {Bertoldi},
  {Chapman}, {Colina}, {Cortes}, {Cox}, {Dickinson}, {Elbaz},
  {G{\'o}nzalez-L{\'o}pez}, {Ibar}, {Infante}, {Hodge}, {Karim}, {Le Fevre},
  {Magnelli}, {Neri}, {Oesch}, {Ota}, {Rix}, {Sargent}, {Sheth}, {van der Wel},
  {van der Werf}, \& {Wagg}}]{decarli+16}
{Decarli}, R., {Walter}, F., {Aravena}, M., {et~al.} 2016, \apj, 833, 69

\bibitem[{{Decarli} {et~al.}(2019){Decarli}, {Walter},
  {G{\'o}nzalez-L{\'o}pez}, {Aravena}, {Boogaard}, {Carilli}, {Cox}, {Daddi},
  {Popping}, {Riechers}, {Uzgil}, {Weiss}, {Assef}, {Bacon}, {Bauer},
  {Bertoldi}, {Bouwens}, {Contini}, {Cortes}, {da Cunha}, {D{\'\i}az-Santos},
  {Elbaz}, {Inami}, {Hodge}, {Ivison}, {Le F{\`e}vre}, {Magnelli}, {Novak},
  {Oesch}, {Rix}, {Sargent}, {Smail}, {Swinbank}, {Somerville}, {van der Werf},
  {Wagg}, \& {Wisotzki}}]{decarli+19}
{Decarli}, R., {Walter}, F., {G{\'o}nzalez-L{\'o}pez}, J., {et~al.} 2019, \apj,
  882, 138

\bibitem[{{Decarli} {et~al.}(2020){Decarli}, {Aravena}, {Boogaard}, {Carilli},
  {Gonz{\'a}lez-L{\'o}pez}, {Walter}, {Cortes}, {Cox}, {da Cunha}, {Daddi},
  {D{\'\i}az-Santos}, {Hodge}, {Inami}, {Neeleman}, {Novak}, {Oesch},
  {Popping}, {Riechers}, {Smail}, {Uzgil}, {van der Werf}, {Wagg}, \&
  {Weiss}}]{decarli+20}
{Decarli}, R., {Aravena}, M., {Boogaard}, L., {et~al.} 2020, \apj, 902, 110

\bibitem[{{Durkalec} {et~al.}(2015){Durkalec}, {Le F{\`e}vre}, {Pollo}, {de la
  Torre}, {Cassata}, {Garilli}, {Le Brun}, {Lemaux}, {Maccagni}, {Pentericci},
  {Tasca}, {Thomas}, {Vanzella}, {Zamorani}, {Zucca}, {Amor{\'\i}n},
  {Bardelli}, {Cassar{\`a}}, {Castellano}, {Cimatti}, {Cucciati}, {Fontana},
  {Giavalisco}, {Grazian}, {Hathi}, {Ilbert}, {Paltani}, {Ribeiro}, {Schaerer},
  {Scodeggio}, {Sommariva}, {Talia}, {Tresse}, {Vergani}, {Capak}, {Charlot},
  {Contini}, {Cuby}, {Dunlop}, {Fotopoulou}, {Koekemoer}, {L{\'o}pez-Sanjuan},
  {Mellier}, {Pforr}, {Salvato}, {Scoville}, {Taniguchi}, \&
  {Wang}}]{durkalec+15}
{Durkalec}, A., {Le F{\`e}vre}, O., {Pollo}, A., {et~al.} 2015, \aap, 583, A128

\bibitem[{{Foreman-Mackey} {et~al.}(2013){Foreman-Mackey}, {Hogg}, {Lang}, \&
  {Goodman}}]{foreman-mackey+13}
{Foreman-Mackey}, D., {Hogg}, D.~W., {Lang}, D., \& {Goodman}, J. 2013, \pasp,
  125, 306

\bibitem[{{Furlanetto} \& {Lidz}(2007)}]{furlanetto+07}
{Furlanetto}, S.~R., \& {Lidz}, A. 2007, \apj, 660, 1030

\bibitem[{{Geach} {et~al.}(2012){Geach}, {Sobral}, {Hickox}, {Wake}, {Smail},
  {Best}, {Baugh}, \& {Stott}}]{geach+12}
{Geach}, J.~E., {Sobral}, D., {Hickox}, R.~C., {et~al.} 2012, \mnras, 426, 679

\bibitem[{{Gong} {et~al.}(2011){Gong}, {Cooray}, {Silva}, {Santos}, \&
  {Lubin}}]{gong+11}
{Gong}, Y., {Cooray}, A., {Silva}, M.~B., {Santos}, M.~G., \& {Lubin}, P. 2011,
  \apjl, 728, L46

\bibitem[{{Gonz{\'a}lez-L{\'o}pez} {et~al.}(2019){Gonz{\'a}lez-L{\'o}pez},
  {Decarli}, {Pavesi}, {Walter}, {Aravena}, {Carilli}, {Boogaard}, {Popping},
  {Weiss}, {Assef}, {Bauer}, {Bertoldi}, {Bouwens}, {Contini}, {Cortes}, {Cox},
  {da Cunha}, {Daddi}, {D{\'\i}az-Santos}, {Inami}, {Hodge}, {Ivison}, {Le
  F{\`e}vre}, {Magnelli}, {esch}, {Riechers}, {Rix}, {Smail}, {Swinbank},
  {Somerville}, {Uzgil}, \& {van der Werf}}]{gonzalez-lopez+19}
{Gonz{\'a}lez-L{\'o}pez}, J., {Decarli}, R., {Pavesi}, R., {et~al.} 2019, arXiv
  e-prints, arXiv:1903.09161

\bibitem[{{Herrero Alonso} {et~al.}(2021){Herrero Alonso}, {Krumpe},
  {Wisotzki}, {Miyaji}, {Garel}, {Schmidt}, {Diener}, {Urrutia}, {Kerutt},
  {Herenz}, {Schaye}, {Pezzulli}, {Maseda}, {Boogaard}, \&
  {Richard}}]{alonso+21_arxiv}
{Herrero Alonso}, Y., {Krumpe}, M., {Wisotzki}, L., {et~al.} 2021, arXiv
  e-prints, arXiv:2107.03723

\bibitem[{{Ihle} {et~al.}(2019){Ihle}, {Chung}, {Stein}, {Alvarez}, {Bond},
  {Breysse}, {Cleary}, {Eriksen}, {Foss}, {Gundersen}, {Harper}, {Murray},
  {Padmanabhan}, {Viero}, {Wehus}, \& {COMAP Collaboration}}]{ihle+19}
{Ihle}, H.~T., {Chung}, D., {Stein}, G., {et~al.} 2019, \apj, 871, 75

\bibitem[{{Inami} {et~al.}(2020){Inami}, {Decarli}, {Walter}, {Weiss},
  {Carilli}, {Aravena}, {Boogaard}, {Gonza{\'l}ez-L{\'o}pez}, {Popping}, {da
  Cunha}, {Bacon}, {Bauer}, {Contini}, {Cortes}, {Cox}, {Daddi},
  {D{\'\i}az-Santos}, {Kaasinen}, {Riechers}, {Wagg}, {van der Werf}, \&
  {Wisotzki}}]{inami+20}
{Inami}, H., {Decarli}, R., {Walter}, F., {et~al.} 2020, \apj, 902, 113

\bibitem[{{Kashikawa} {et~al.}(2006){Kashikawa}, {Yoshida}, {Shimasaku},
  {Nagashima}, {Yahagi}, {Ouchi}, {Matsuda}, {Malkan}, {Doi}, {Iye}, {Ajiki},
  {Akiyama}, {Ando}, {Aoki}, {Furusawa}, {Hayashino}, {Iwamuro}, {Karoji},
  {Kobayashi}, {Kodaira}, {Kodama}, {Komiyama}, {Miyazaki}, {Mizumoto},
  {Morokuma}, {Motohara}, {Murayama}, {Nagao}, {Nariai}, {Ohta}, {Okamura},
  {Sasaki}, {Sato}, {Sekiguchi}, {Shioya}, {Tamura}, {Taniguchi}, {Umemura},
  {Yamada}, \& {Yasuda}}]{kashikawa+06}
{Kashikawa}, N., {Yoshida}, M., {Shimasaku}, K., {et~al.} 2006, \apj, 637, 631

\bibitem[{{Keating} {et~al.}(2020){Keating}, {Marrone}, {Bower}, \&
  {Keenan}}]{keating+20}
{Keating}, G.~K., {Marrone}, D.~P., {Bower}, G.~C., \& {Keenan}, R.~P. 2020,
  \apj, 901, 141

\bibitem[{{Keating} {et~al.}(2016){Keating}, {Marrone}, {Bower}, {Leitch},
  {Carlstrom}, \& {DeBoer}}]{keating+16}
{Keating}, G.~K., {Marrone}, D.~P., {Bower}, G.~C., {et~al.} 2016, \apj, 830,
  34

\bibitem[{{Keating} {et~al.}(2015){Keating}, {Bower}, {Marrone}, {DeBoer},
  {Heiles}, {Chang}, {Carlstrom}, {Greer}, {Hawkins}, {Lamb}, {Leitch},
  {Miller}, {Muchovej}, \& {Woody}}]{keating+15}
{Keating}, G.~K., {Bower}, G.~C., {Marrone}, D.~P., {et~al.} 2015, \apj, 814,
  140

\bibitem[{{Keenan} {et~al.}(2020){Keenan}, {Marrone}, \& {Keating}}]{keenan+20}
{Keenan}, R.~P., {Marrone}, D.~P., \& {Keating}, G.~K. 2020, \apj, 904, 127

\bibitem[{{Kennicutt}(1998)}]{kennicutt98}
{Kennicutt}, Robert~C., J. 1998, \apj, 498, 541

\bibitem[{{Khostovan} {et~al.}(2019){Khostovan}, {Sobral}, {Mobasher},
  {Matthee}, {Cochrane}, {Chartab}, {Jafariyazani}, {Paulino-Afonso}, {Santos},
  \& {Calhau}}]{khostovan+19}
{Khostovan}, A.~A., {Sobral}, D., {Mobasher}, B., {et~al.} 2019, \mnras, 489,
  555

\bibitem[{{Kriek} {et~al.}(2015){Kriek}, {Shapley}, {Reddy}, {Siana}, {Coil},
  {Mobasher}, {Freeman}, {de Groot}, {Price}, {Sanders}, {Shivaei}, {Brammer},
  {Momcheva}, {Skelton}, {van Dokkum}, {Whitaker}, {Aird}, {Azadi}, {Kassis},
  {Bullock}, {Conroy}, {Dav{\'e}}, {Kere{\v s}}, \& {Krumholz}}]{kriek+15}
{Kriek}, M., {Shapley}, A.~E., {Reddy}, N.~A., {et~al.} 2015, \apjs, 218, 15

\bibitem[{{Lenki{\'c}} {et~al.}(2020){Lenki{\'c}}, {Bolatto}, {F{\"o}rster
  Schreiber}, {Tacconi}, {Neri}, {Combes}, {Walter}, {Garc{\'\i}a-Burillo},
  {Genzel}, {Lutz}, \& {Cooper}}]{lenkic+20}
{Lenki{\'c}}, L., {Bolatto}, A.~D., {F{\"o}rster Schreiber}, N.~M., {et~al.}
  2020, \aj, 159, 190

\bibitem[{{Li} {et~al.}(2016){Li}, {Wechsler}, {Devaraj}, \& {Church}}]{li+16}
{Li}, T.~Y., {Wechsler}, R.~H., {Devaraj}, K., \& {Church}, S.~E. 2016, \apj,
  817, 169

\bibitem[{{Lidz} {et~al.}(2011){Lidz}, {Furlanetto}, {Oh}, {Aguirre}, {Chang},
  {Dor{\'e}}, \& {Pritchard}}]{lidz+11}
{Lidz}, A., {Furlanetto}, S.~R., {Oh}, S.~P., {et~al.} 2011, \apj, 741, 70

\bibitem[{{Marinacci} {et~al.}(2018){Marinacci}, {Vogelsberger}, {Pakmor},
  {Torrey}, {Springel}, {Hernquist}, {Nelson}, {Weinberger}, {Pillepich},
  {Naiman}, \& {Genel}}]{TNG4}
{Marinacci}, F., {Vogelsberger}, M., {Pakmor}, R., {et~al.} 2018, \mnras, 480,
  5113

\bibitem[{{Masui} {et~al.}(2013){Masui}, {Switzer}, {Banavar}, {Bandura},
  {Blake}, {Calin}, {Chang}, {Chen}, {Li}, {Liao}, {Natarajan}, {Pen},
  {Peterson}, {Shaw}, \& {Voytek}}]{masui+13}
{Masui}, K.~W., {Switzer}, E.~R., {Banavar}, N., {et~al.} 2013, \apjl, 763, L20

\bibitem[{{Momcheva} {et~al.}(2016){Momcheva}, {Brammer}, {van Dokkum},
  {Skelton}, {Whitaker}, {Nelson}, {Fumagalli}, {Maseda}, {Leja}, {Franx},
  {Rix}, {Bezanson}, {Da Cunha}, {Dickey}, {F{\"o}rster Schreiber},
  {Illingworth}, {Kriek}, {Labb{\'e}}, {Ulf Lange}, {Lundgren}, {Magee},
  {Marchesini}, {Oesch}, {Pacifici}, {Patel}, {Price}, {Tal}, {Wake}, {van der
  Wel}, \& {Wuyts}}]{momcheva+16}
{Momcheva}, I.~G., {Brammer}, G.~B., {van Dokkum}, P.~G., {et~al.} 2016, \apjs,
  225, 27

\bibitem[{{Moradinezhad Dizgah} \& {Keating}(2019)}]{dizgah+19}
{Moradinezhad Dizgah}, A., \& {Keating}, G.~K. 2019, \apj, 872, 126

\bibitem[{{Murray} {et~al.}(2020){Murray}, {Diemer}, \& {Chen}}]{halomod}
{Murray}, S.~G., {Diemer}, B., \& {Chen}, Z. 2020, arXiv e-prints,
  arXiv:2009.14066

\bibitem[{{Naiman} {et~al.}(2018){Naiman}, {Pillepich}, {Springel},
  {Ramirez-Ruiz}, {Torrey}, {Vogelsberger}, {Pakmor}, {Nelson}, {Marinacci},
  {Hernquist}, {Weinberger}, \& {Genel}}]{TNG2}
{Naiman}, J.~P., {Pillepich}, A., {Springel}, V., {et~al.} 2018, \mnras, 477,
  1206

\bibitem[{{Nelson} {et~al.}(2018){Nelson}, {Pillepich}, {Springel},
  {Weinberger}, {Hernquist}, {Pakmor}, {Genel}, {Torrey}, {Vogelsberger},
  {Kauffmann}, {Marinacci}, \& {Naiman}}]{TNG1}
{Nelson}, D., {Pillepich}, A., {Springel}, V., {et~al.} 2018, \mnras, 475, 624

\bibitem[{{Oxholm} \& {Switzer}(2021)}]{oxholm+21_arxiv}
{Oxholm}, T.~M., \& {Switzer}, E.~R. 2021, arXiv e-prints, arXiv:2107.02111

\bibitem[{{Pavesi} {et~al.}(2018){Pavesi}, {Sharon}, {Riechers}, {Hodge},
  {Decarli}, {Walter}, {Carilli}, {Daddi}, {Smail}, {Dickinson}, {Ivison},
  {Sargent}, {da Cunha}, {Aravena}, {Darling}, {Smol{\v c}i{\'c}}, {Scoville},
  {Capak}, \& {Wagg}}]{pavesi+18}
{Pavesi}, R., {Sharon}, C.~E., {Riechers}, D.~A., {et~al.} 2018, \apj, 864, 49

\bibitem[{{Pillepich} {et~al.}(2018){Pillepich}, {Nelson}, {Hernquist},
  {Springel}, {Pakmor}, {Torrey}, {Weinberger}, {Genel}, {Naiman}, {Marinacci},
  \& {Vogelsberger}}]{TNG3}
{Pillepich}, A., {Nelson}, D., {Hernquist}, L., {et~al.} 2018, \mnras, 475, 648

\bibitem[{{Pullen} {et~al.}(2013){Pullen}, {Chang}, {Dor{\'e}}, \&
  {Lidz}}]{pullen+13}
{Pullen}, A.~R., {Chang}, T.-C., {Dor{\'e}}, O., \& {Lidz}, A. 2013, \apj, 768,
  15

\bibitem[{{Reddy} {et~al.}(2006){Reddy}, {Steidel}, {Erb}, {Shapley}, \&
  {Pettini}}]{reddy+06}
{Reddy}, N.~A., {Steidel}, C.~C., {Erb}, D.~K., {Shapley}, A.~E., \& {Pettini},
  M. 2006, \apj, 653, 1004

\bibitem[{{Riechers} {et~al.}(2019){Riechers}, {Pavesi}, {Sharon}, {Hodge},
  {Decarli}, {Walter}, {Carilli}, {Aravena}, {da Cunha}, {Daddi}, {Dickinson},
  {Smail}, {Capak}, {Ivison}, {Sargent}, {Scoville}, \& {Wagg}}]{riechers+18}
{Riechers}, D.~A., {Pavesi}, R., {Sharon}, C.~E., {et~al.} 2019, \apj, 872, 7

\bibitem[{{Silva} {et~al.}(2015){Silva}, {Santos}, {Cooray}, \&
  {Gong}}]{silva+15}
{Silva}, M., {Santos}, M.~G., {Cooray}, A., \& {Gong}, Y. 2015, \apj, 806, 209

\bibitem[{{Skelton} {et~al.}(2014){Skelton}, {Whitaker}, {Momcheva}, {Brammer},
  {van Dokkum}, {Labb{\'e}}, {Franx}, {van der Wel}, {Bezanson}, {Da Cunha},
  {Fumagalli}, {F{\"o}rster Schreiber}, {Kriek}, {Leja}, {Lundgren}, {Magee},
  {Marchesini}, {Maseda}, {Nelson}, {Oesch}, {Pacifici}, {Patel}, {Price},
  {Rix}, {Tal}, {Wake}, \& {Wuyts}}]{skelton+14}
{Skelton}, R.~E., {Whitaker}, K.~E., {Momcheva}, I.~G., {et~al.} 2014, \apjs,
  214, 24

\bibitem[{{Springel} {et~al.}(2018){Springel}, {Pakmor}, {Pillepich},
  {Weinberger}, {Nelson}, {Hernquist}, {Vogelsberger}, {Genel}, {Torrey},
  {Marinacci}, \& {Naiman}}]{TNG5}
{Springel}, V., {Pakmor}, R., {Pillepich}, A., {et~al.} 2018, \mnras, 475, 676

\bibitem[{{Steidel} {et~al.}(2004){Steidel}, {Shapley}, {Pettini},
  {Adelberger}, {Erb}, {Reddy}, \& {Hunt}}]{steidel+04}
{Steidel}, C.~C., {Shapley}, A.~E., {Pettini}, M., {et~al.} 2004, \apj, 604,
  534

\bibitem[{{Sun} {et~al.}(2019){Sun}, {Hensley}, {Chang}, {Dor{\'e}}, \&
  {Serra}}]{sun+19}
{Sun}, G., {Hensley}, B.~S., {Chang}, T.-C., {Dor{\'e}}, O., \& {Serra}, P.
  2019, \apj, 887, 142

\bibitem[{{Sun} {et~al.}(2021){Sun}, {Chang}, {Uzgil}, {Bock}, {Bradford},
  {Butler}, {Caze-Cortes}, {Cheng}, {Cooray}, {Crites}, {Hailey-Dunsheath},
  {Emerson}, {Frez}, {Hoscheit}, {Hunacek}, {Keenan}, {Li}, {Madonia},
  {Marrone}, {Moncelsi}, {Shiu}, {Trumper}, {Turner}, {Weber}, {Wei}, \&
  {Zemcov}}]{sun+21}
{Sun}, G., {Chang}, T.~C., {Uzgil}, B.~D., {et~al.} 2021, \apj, 915, 33

\bibitem[{{Tinker} {et~al.}(2008){Tinker}, {Kravtsov}, {Klypin}, {Abazajian},
  {Warren}, {Yepes}, {Gottl{\"o}ber}, \& {Holz}}]{tinker+08}
{Tinker}, J., {Kravtsov}, A.~V., {Klypin}, A., {et~al.} 2008, \apj, 688, 709

\bibitem[{{Tinker} {et~al.}(2010){Tinker}, {Robertson}, {Kravtsov}, {Klypin},
  {Warren}, {Yepes}, \& {Gottl{\"o}ber}}]{tinker+10}
{Tinker}, J.~L., {Robertson}, B.~E., {Kravtsov}, A.~V., {et~al.} 2010, \apj,
  724, 878

\bibitem[{{Uzgil} {et~al.}(2019){Uzgil}, {Carilli}, {Lidz}, {Walter},
  {Thyagarajan}, {Decarli}, {Aravena}, {Bertoldi}, {Cortes},
  {Gonz{\'a}lez-L{\'o}pez}, {Inami}, {Popping}, {Riechers}, {Van der Werf},
  {Wagg}, \& {Weiss}}]{uzgil+19}
{Uzgil}, B.~D., {Carilli}, C., {Lidz}, A., {et~al.} 2019, \apj, 887, 37

\bibitem[{{Visbal} \& {Loeb}(2010)}]{visbal+10}
{Visbal}, E., \& {Loeb}, A. 2010, \jcap, 2010, 016

\bibitem[{{Walter} {et~al.}(2014){Walter}, {Decarli}, {Sargent}, {Carilli},
  {Dickinson}, {Riechers}, {Ellis}, {Stark}, {Weiner}, {Aravena}, {Bell},
  {Bertoldi}, {Cox}, {Da Cunha}, {Daddi}, {Downes}, {Lentati}, {Maiolino},
  {Menten}, {Neri}, {Rix}, \& {Weiss}}]{walter+14}
{Walter}, F., {Decarli}, R., {Sargent}, M., {et~al.} 2014, \apj, 782, 79

\bibitem[{{Walter} {et~al.}(2020){Walter}, {Carilli}, {Neeleman}, {Decarli},
  {Popping}, {Somerville}, {Aravena}, {Bertoldi}, {Boogaard}, {Cox}, {da
  Cunha}, {Magnelli}, {Obreschkow}, {Riechers}, {Rix}, {Smail}, {Weiss},
  {Assef}, {Bauer}, {Bouwens}, {Contini}, {Cortes}, {Daddi}, {Diaz-Santos},
  {Gonz{\'a}lez-L{\'o}pez}, {Hennawi}, {Hodge}, {Inami}, {Ivison}, {Oesch},
  {Sargent}, {van der Werf}, {Wagg}, \& {Yung}}]{walter+20}
{Walter}, F., {Carilli}, C., {Neeleman}, M., {et~al.} 2020, \apj, 902, 111

\bibitem[{{Wirth} {et~al.}(2015){Wirth}, {Trump}, {Barro}, {Guo}, {Koo}, {Liu},
  {Kassis}, {Lyke}, {Rizzi}, {Campbell}, {Goodrich}, \& {Faber}}]{wirth+15}
{Wirth}, G.~D., {Trump}, J.~R., {Barro}, G., {et~al.} 2015, \aj, 150, 153

\bibitem[{{Wolz} {et~al.}(2017){Wolz}, {Blake}, \& {Wyithe}}]{wolz+17}
{Wolz}, L., {Blake}, C., \& {Wyithe}, J.~S.~B. 2017, \mnras, 470, 3220

\bibitem[{{Wolz} {et~al.}(2016){Wolz}, {Tonini}, {Blake}, \&
  {Wyithe}}]{wolz+16}
{Wolz}, L., {Tonini}, C., {Blake}, C., \& {Wyithe}, J.~S.~B. 2016, \mnras, 458,
  3399

\bibitem[{{Wolz} {et~al.}(2021){Wolz}, {Pourtsidou}, {Masui}, {Chang},
  {Bautista}, {Mueller}, {Avila}, {Bacon}, {Percival}, {Cunnington},
  {Anderson}, {Chen}, {Kneib}, {Li}, {Liao}, {Pen}, {Peterson}, {Rossi},
  {Schneider}, {Yadav}, \& {Zhao}}]{wolz+21_arxiv}
{Wolz}, L., {Pourtsidou}, A., {Masui}, K.~W., {et~al.} 2021, arXiv e-prints,
  arXiv:2102.04946

\bibitem[{{Yang} {et~al.}(2021{\natexlab{a}}){Yang}, {Popping}, {Somerville},
  {Pullen}, {Breysse}, \& {Maniyar}}]{yang+21b_arxiv}
{Yang}, S., {Popping}, G., {Somerville}, R.~S., {et~al.} 2021{\natexlab{a}},
  arXiv e-prints, arXiv:2108.07716

\bibitem[{{Yang} {et~al.}(2021{\natexlab{b}}){Yang}, {Somerville}, {Pullen},
  {Popping}, {Breysse}, \& {Maniyar}}]{yang+21a}
{Yang}, S., {Somerville}, R.~S., {Pullen}, A.~R., {et~al.} 2021{\natexlab{b}},
  \apj, 911, 132

\bibitem[{{Yoshikawa} {et~al.}(2010){Yoshikawa}, {Akiyama}, {Kajisawa},
  {Alexander}, {Ohta}, {Suzuki}, {Tokoku}, {Uchimoto}, {Konishi}, {Yamada},
  {Tanaka}, {Omata}, {Nishimura}, {Koekemoer}, {Brandt}, \&
  {Ichikawa}}]{yoshikawa+10}
{Yoshikawa}, T., {Akiyama}, M., {Kajisawa}, M., {et~al.} 2010, \apj, 718, 112

\end{thebibliography}

\end{document}